\documentclass{smjour}
\usepackage{blank}
\usepackage{makeidx}
\usepackage{amsfonts}
\usepackage{amsmath}
\usepackage{amssymb}
\usepackage{amsbsy}
\usepackage[dvips]{graphicx}
\usepackage{algorithmic}
\usepackage{algorithm}
\usepackage{subfigure}
\usepackage{enumerate}
\usepackage{bm}
\usepackage{simplemargins}
\usepackage{xspace}
\setallmargins{0.95in}

\newenvironment{thebibliography}[1]{\begin{references}}{\end{references}}

\newtheorem{question}{Question}[question]

\let\doendproof\endproof
\def\endproof{\doendproof\medskip}

\def\O{\mathcal{O}{}}
\def\short{\mathfrak{s}}

\newcommand{\heading}[1]{\smallskip\par\noindent{\bf #1}}
\def\ACCEPT{{\sffamily ACCEPT}\xspace}
\def\REJECT{{\sffamily REJECT}\xspace}

  \def\calC{{\cal C}}

 \def\calR{{\cal R}}

\newenvironment{packed_enum}{
	\begin{enumerate}
		\setlength{\itemsep}{1pt}
	    \setlength{\parskip}{0pt}
		\setlength{\parsep}{0pt}
}{\end{enumerate}}

\newenvironment{packed_itemize}{
	\begin{itemize}
		\setlength{\itemsep}{1pt}
	    \setlength{\parskip}{0pt}
		\setlength{\parsep}{0pt}
}{\end{itemize}}

\newenvironment{packed_head_enum}[1]{
	\begin{enumerate}[#1]
		\setlength{\itemsep}{1pt}
	    \setlength{\parskip}{0pt}
		\setlength{\parsep}{0pt}
}{\end{enumerate}}

\def\computationproblem#1#2#3#4{
	\begin{center}
	{\begin{tabular}{rp{#4}}
	{\bf Problem:}&#1\\
	{\bf Input:}&#2\\
	{\bf Output:}&#3\\
	\end{tabular}}
	\end{center}
	\medskip
}

\def\lftcrossed{\mathrel{\kern 0.15em\uparrow\kern -0.90em\raisebox{-0.15em}{$\rightarrow$}}}
\def\rtcrossed{\mathrel{\kern 0.4em\uparrow\kern -1.2em\raisebox{-0.15em}{$\leftarrow$}}}

\def\fourrel{\mathrel{\bigcirc\kern -0.905em \raisebox{0.075em}{$\scriptstyle{\uparrow\downarrow}$}}}
\def\notfourrel{\kern 0.1em\not\kern -0.1em\fourrel}

\def\cNP{\hbox{\rm \sffamily NP}}
\def\cFPT{\hbox{\rm \sffamily FPT}}

\def\int{\hbox{\bf \rm \sffamily INT}}

\def\path{\hbox{\bf \rm \sffamily PATH}}

\def\circle{\hbox{\bf \rm \sffamily CIRCLE}}
\def\perm{\hbox{\bf \rm \sffamily PERM}}

\def\ext{\textsc{RepExt}}
\def\simrep{\textsc{Sim}}
\def\totalorder{\textsc{TotalOrdering}}

\begin{document}

\title{Extending Partial Representations\\of Circle Graphs\thanks{%
The conference version of this paper appeared in Graph Drawing 2013~\cite{cfk}.}}

\titlerunninghead{Extending Partial Representations of Circle Graphs}
\author{Steven Chaplick}
\affil{Lehrstuhl f\"ur Informatik I, Universit\"at W\"urzburg, Germany.\\
E-mail: \texttt{steven.chaplick@uni-wuerzburg.de}.}

\authors{Radoslav Fulek}
\affil{Institute of Science and Technology Austria, Klosterneuburg, Austria. E-mail:
\texttt{radoslav.fulek@gmail.com}.}

\authors{Pavel Klav\'ik}
\affil{Computer Science Institute, Charles University in Prague,\\Czech Republic.
E-mail: \texttt{klavik@iuuk.mff.cuni.cz}.}

\abstract{\emph{The partial representation extension problem} is a recently introduced
generalization of the recognition problem. A \emph{circle graph} is an intersection graph of chords
of a circle.  We study the partial representation extension problem for circle graphs, where the
input consists of a graph $G$ and a partial representation $\calR'$ giving some pre-drawn chords
that represent an induced subgraph of $G$. The question is whether one can extend $\calR'$ to a
representation $\calR$ of the entire graph $G$, i.e., whether one can draw the remaining chords into
a partially pre-drawn representation to obtain a representation of $G$. Our main result is an
$\O(n^3)$ time algorithm for partial representation extension of circle graphs, where $n$ is the
number of vertices. To show this, we describe the structure of all representations of a circle graph
using split decomposition. This can be of independent interest.}

\begin{article}

\section{Introduction} \label{sec:introduction}

Geometric graph representations are important topics of graph theory and computer science.  A
frequently studied type of representations are the so-called \emph{intersection representations}. An
intersection representation of a graph represents its vertices by some objects and encodes its edges
by intersections of these objects, i.e., two vertices are adjacent if and only if the corresponding
objects intersect. Classes of intersection graphs are obtained by restricting these objects; e.g.,
\emph{interval graphs} are intersection graphs of intervals of the real line, \emph{string graphs}
are intersection graphs of curves in plane, and so on. These representations are well-studied; see
e.g.~\cite{egr}.

\begin{figure}[t!]
\centering
\includegraphics{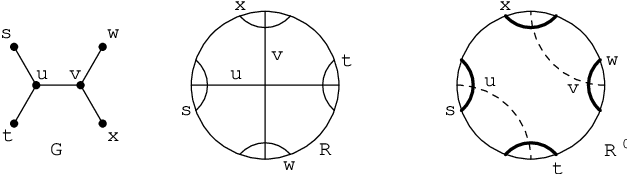}
\caption{On the left, a circle graph $G$ with a representation $\calR$ is given. A partial
representation $\calR'$ given on the right with the pre-drawn chords $\boldsymbol s$, $\boldsymbol
t$, $\boldsymbol w$, and $\boldsymbol x$ is not extendible. The chords are depicted as arcs to make
the figure more readable.}
\label{fig:partial_representation_example}
\end{figure}

For a fixed class $\calC$ of intersection-defined graphs, a very natural computational problem is
\emph{recognition}. It asks whether an input graph $G$ belongs to $\calC$.  In this paper, we study
a recently introduced generalization of this problem called \emph{partial representation
extension}~\cite{KKV}. Its input gives with $G$ a part of the representation and the problem asks
whether this partial representation can be extended to a representation of the entire $G$;
see Fig.~\ref{fig:partial_representation_example} for an illustration. We show
that this problem can be solved in polynomial time for the class of \emph{circle graphs}.

\begin{figure}[b!]
\centering
\includegraphics{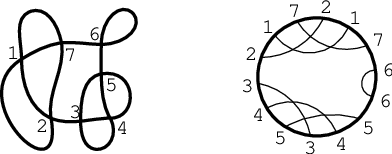}
\caption{A self-intersecting closed curve with $n$ intersections numbered $1,\dots,n$ corresponds to
a representation of circle graph with the vertices $1,\dots,n$ where the endpoints of the chords are
placed according to the order of the intersections along the curve.}
\label{fig:gauss_words}
\end{figure}

\heading{Circle Graphs.}
Circle graphs are intersection graphs of chords of a circle. They were first considered by Even and
Itai~\cite{EI71} in the early 1970s in study of stack sorting techniques. Other motivations are due
to their relations to Gauss words~\cite{deFM99} (see Fig.~\ref{fig:gauss_words}) and matroid
representations~\cite{deF81,B87}. Circle graphs are also important regarding rank-width~\cite{oum}.

Let $\chi(G)$ denote the chromatic number of $G$, and let $\omega(G)$ denote the clique-number of $G$.
Trivially we have $\omega(G) \le \chi(G)$ and the graphs for which every induced subgraph satisfies
equality are the well-known \emph{perfect graphs}~\cite{spgt}. In general, the difference between these
two numbers can be arbitrarily high, e.g., there is a triangle-free graph with an arbitrary high chromatic
number. Circle graphs are known to be \emph{almost perfect} which means that $\chi(G) \le
f(\omega(G))$ for some function $f$. The best known result for circle graphs~\cite{KK97} states that
$f(k)$ is $\Omega(k \log k)$ and $\O(2^k)$.

Some \cNP-hard problems, such as maximum weighted clique
and independent set~\cite{G00}, become tractable on circle graphs. On the other hand, problems
such as vertex colorability~\cite{GJMP80} and Hamiltonicity~\cite{D89} remain \cNP-complete even for
circle graphs.

The complexity of recognition of circle graphs was a long standing open problem; see~\cite{egr} for an
overview. The first results, e.g.,~\cite{EI71}, gave existential characterizations which did not give
polynomial-time algorithms. The mystery whether circle graphs can be recognized in polynomial time
frustrated mathematicians for some years. It was resolved in the mid-1980s and several
polynomial-time algorithms were discovered~\cite{B1987,GSH89,N85} (in time $\O(n^7)$ and similar).
Later, a more efficient algorithm~\cite{S94} based on \emph{split decomposition} was given, and
the current state-of-the-art recognition algorithm~\cite{GPTC13} runs in a quasi-linear time in the
number of vertices and the number of edges of the graph.

\heading{The Partial Representation Extension Problem.}
It is quite surprising that this very natural generalization of the recognition problem was
considered only recently. It is currently an active area of research which is inspiring a deeper
investigation of many classical graph classes.  For instance, a recent result of Angelini et
al.~\cite{ABFJKPR10} states that the problem is decidable in linear time for planar graphs. On the
other hand, F\'ary's Theorem claims that every planar graph has a straight-line embedding, but
extension of such an embedding is \cNP-hard~\cite{patrignani}.

In the context of intersection-defined classes, this problem was first considered in~\cite{KKV}
for interval graphs. Currently, the best known results are linear-time algorithms for interval
graphs~\cite{blas_rutter,KKOSV} and proper interval graphs~\cite{KKORSSV}, a quadratic-time algorithm
for unit interval graphs~\cite{KKORSSV,soulignac,soulignac2}, and polynomial-time algorithms for
permutation and function graphs~\cite{KKKW}, proper circular-arc graphs~\cite{jensen}, and trapezoid
graphs~\cite{trapezoid_repext}. For chordal graphs (as subtree-in-a-tree graphs) several versions of
the problems were considered~\cite{KKOS} and all of them are \cNP-complete, and similarly for
different contact representations of planar graphs~\cite{contact_planar_ext}.  In~\cite{KS}, minimal
forbidden configurations making a partial interval representation non-extendible are characterized.
Extending partial visibility representations is studied in~\cite{ext_visibility}.

\heading{The Structure of Representations.}
To solve the recognition problem for $G$, one just needs to build a single representation.
However, to solve the partial representation extension problem, the structure of all
representations of $G$ must be well understood.  A general approach used in the above papers is the
following. We first derive necessary and sufficient constraints from the partial representation
$\calR'$. Then we efficiently test whether some representation $\calR$ satisfies these constraints.
If none satisfies them, then $\calR'$ is not extendible. And if some $\calR$ satisfies them, then it
extends $\calR'$.

It is well-known that the split decomposition~\cite[Theorem 3]{Cunningham82} captures the structure
of all representations of circle graphs.  The standard recognition algorithms produce a special type
of representations using split decomposition as follows. We find a \emph{split} in $G$, construct
two smaller graphs, build their representation recursively, and then join these two representations
to produce $\calR$.  In Section~\ref{sec:structural_results},  we give a simple recursive
description of all possible representations based on splits. Our result can be interpreted as
``describing a structure like PQ-trees\footnote{See~\cite{PQ_trees} for further information on PQ-trees.} 
for circle graphs.'' It is possible that the proof techniques
from other papers on circle graphs such as~\cite{Courcelle08,GPTC13} would give a similar
description. However, these techniques are more involved than our approach which turns out to be
quite elementary and simple.

\heading{Restricted Representations.} The partial representation extension problem belongs to a
larger group of problems dealing with \emph{restricted representations of graphs}. These problems
ask whether there is some representation of an input graph $G$ satisfying some additional
constraints. We describe two examples of these problems.

An input of the \emph{simultaneous representation problem}\footnote{Here, we will focus on what is
sometimes referred to as the \emph{sunflower version} in the literature, see~\cite{SimRep_Planar}.},
shortly $\simrep$, consists of graphs $G_1,\dots,G_k$ with some vertices common for all the graphs.
The problem asks whether there exist representations $\calR_1,\dots,\calR_k$ representing the common
vertices in the same way.  This problem is polynomially solvable for permutation and comparability
graphs~\cite{simrep}. They additionally show that for chordal graphs it is \cNP-complete when $k$ is
part of the input and polynomially solvable for $k = 2$.  For interval graphs, a linear-time
algorithm is known for $k=2$~\cite{blas_rutter} and the complexity is open in general. For some
classes, these problems are closely related to the partial representation extension problems.  For
example, there is an \cFPT\ algorithm for interval graphs with the number of common vertices as the
parameter~\cite{KKV}, and partial representations of interval graphs can be extended in linear time
by reducing it to corresponding simultaneous representation problem~\cite{blas_rutter}.

The \emph{bounded representation problem}~\cite{KKORSSV} prescribes bounds for each vertex of the
input graph and asks whether there is some representation satisfying these bounds. For circle
graphs, the input specifies for each chord $v$ a pair of arcs $(A_v,A'_v)$ of the circle, and a
solution is required to have one endpoint of $v$ in $A_v$ and the other one in $A'_v$. This problem
is clearly a generalization of partial representation extension since one can describe a partial
representation using singleton arcs.  It is known to be polynomially solvable for interval and
proper interval representations of interval graphs~\cite{BKO}, and surprisingly it is \cNP-complete
for unit interval representations~\cite{KKORSSV,soulignac,soulignac2}. The complexity for other
classes is not known.

\heading{Our Results.} We study the following problem (see Section~\ref{sec:preliminaries} for
definitions):

\computationproblem
{Partial Representation Extension -- $\ext(\circle)$}
{A circle graph $G$ and a partial representation $\calR'$.}
{Is there a representation $\calR$ of $G$ extending $\calR'$?}
{8.5cm}

In Section~\ref{sec:structural_results}, we describe a simple structure of all representations. This
is used in Section~\ref{sec:algorithm} to obtain our main algorithmic result:

\begin{theorem} \label{thm:ext_circle}
The problem\/ $\ext(\circle)$ can be solved in time $\O(n^3)$ where $n$ is the number of vertices.
\end{theorem}
\bigskip
To spice up our results, we show in Section~\ref{sec:simultaneous_representations} the following for
the simultaneous representation problem of circle graphs:

\begin{theorem} \label{prop:simrep_npc}
If $k$ is a part of the input, the problem\/ $\simrep(\circle)$ of $k$ circle graphs is\/ \cNP-complete.
\end{theorem}

Finally, we show that Theorem~\ref{thm:ext_circle} implies the following.

\begin{corollary} \label{cor:simrep_fpt}
The problem\/ $\simrep(\circle)$ is\/ \cFPT\ in the size of the common subgraph.
\end{corollary}

\section{Definitions and Preliminaries} \label{sec:preliminaries}

\heading{Circle Representations.} A \emph{circle representation} $\calR$
of  a graph $G$ is a collection $\bigl\{C_u
\mid u \in V(G)\bigr\}$ of chords of a circle such that $C_u$ intersects $C_v$ if and only if $uv
\in E(G)$. A graph is a \emph{circle graph} if it has a circle representation, and we denote the
class of circle graphs by \circle.

Notice that a representation of a circle graph is completely determined by the circular order of
the endpoints of the chords in the representation, and two chords $C_u$ and $C_v$ cross if and only
if their endpoints alternate in this order.  For convenience we label both endpoints of the chord
representing a vertex by the same label as the vertex.

\begin{figure}[t!]
\centering
\includegraphics{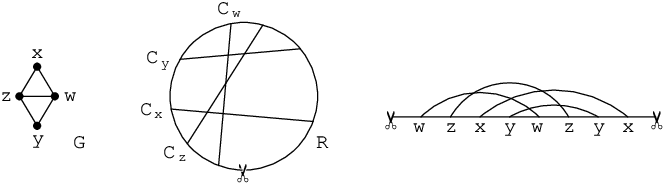}
\caption{An example of a circle graph with a circle graph representation on the left; an interval overlap
representation of the same graph on the right.}
\label{fig:example_of_circle_graph}
\end{figure}

\heading{Interval Overlap Graphs.} Suppose that we cut the circle in a point which is not an
endpoint of a chord and straighten it into a segment; see Fig.~\ref{fig:example_of_circle_graph}.
From this straightening of the circle, each chord can now be seen as an arc above the resulting
segment.  Notice that two chords $C_u$ and $C_v$ cross if and only if their endpoints appear in the
order $uvuv$ or $vuvu$ from left to right. Alternatively, circle graphs are called \emph{interval
overlap graphs}. Their vertices can be represented by intervals and two vertices are adjacent if and
only if their intervals overlap which means they intersect and one is not a subset of the other.

\heading{Word representations.}
A sequence $\tau$ over an alphabet of symbols $\Sigma$ is a \emph{word}.  A \emph{circular word}
represents the set of words which are cyclical shifts of one another. In the sequel, we represent a
circular word by a word from its corresponding set of words.  We denote words and circular words by
small Greek letters.

For a word $\tau$ and a symbol $u$ we write $u \in \tau$, if $u$ appears at least once in $\tau$.
Thus, $\tau$ is also used to denote the set of symbols occurring in $\tau$.  A word $\tau$ is a
\emph{subword} of $\sigma$, if $\tau$ appears consecutively in $\sigma$. A word $\tau$ is a
\emph{subsequence} of $\sigma$, if the word $\tau$ can be obtained from $\sigma$ by deleting some
symbols.  We say that $u$ \emph{alternates} with $v$ in $\tau$, if $uvuv$ or $vuvu$ is a subsequence
of $\tau$.  The corresponding definitions also apply to circular words. If $\sigma$ and $\tau$ are
two words, we denote their concatenation by $\sigma \tau$.

The above interpretation of circle graphs as interval overlap graphs allows us to associate each
representation $\calR$ of $G$ with a unique circular word $\tau$ over $V$.  The word $\tau$ is
obtained by the circular order of the endpoints of the chords in $\calR$ as they appear along the
circle when traversed clockwise.  The occurrences of $u$ and $v$ alternate in $\tau$ if and only if
$uv \in E(G)$.  For example $\calR$ in Fig.~\ref{fig:partial_representation_example} corresponds to
the circular word $\tau = susxvxtutwvw$. Notice that each vertex appears exactly twice in $\tau$.  A
circular subsequence $\tau'$ of $\tau$ is \emph{induced} by $V'\subseteq V(G)$ if $\tau'$ is
obtained from $\tau$ by deleting symbols in $V(G)\setminus V'$. 

\heading{Partial Representations.}
Partial representations are defined in~\cite{KKV} and other papers as representations of induced
subgraphs. In this paper, we consider the following more general definition.  A \emph{partial
representation} $\calR'$ of a circle graph $G$ is given by a circular word $\tau'$ consisting of
symbols of $V(G)$ such that each $u \in V(G)$ appears at most twice in $\tau'$.  A representation
$\calR$ of $G$ corresponding to a circular word $\tau$ \emph{extends} $\calR'$ if and only if
$\tau'$ is a subsequence of $\tau$. The endpoints in $\tau'$ and the corresponding vertices are
called \emph{pre-drawn}. If a pre-drawn vertex $u$ has both occurrences in $\tau'$, the chord $C_u$
is \emph{pre-drawn}.

\section{Structure of Representations of Maximal Splits} \label{sec:structural_results}

\begin{figure}[t!]
\centering
\includegraphics{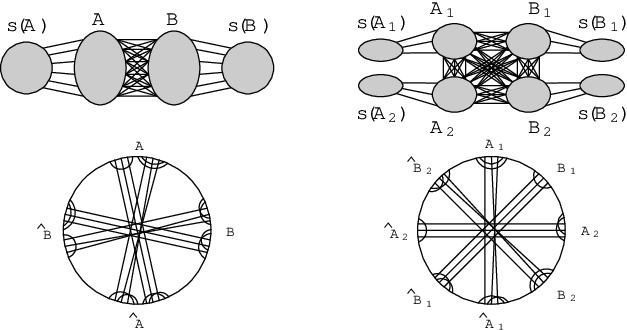}
\caption{Two different representations of $G$ with the split between $A$ and $B$.}
\label{fig:split_representations}
\end{figure}

Let $G$ be a connected graph.  A \emph{split} of $G$ is a partition of the vertices of $G$ into four
parts $A$, $B$, $\short(A)$ and $\short(B)$, such that:
\begin{packed_itemize}
\item We have $A \ne \emptyset$ and $B \ne \emptyset$, but possibly $\short(A) = \emptyset$ or
$\short(B) = \emptyset$.
\item For every $a \in A$ and $b \in B$, we have $ab \in E(G)$.
\item There is no edge between $\short(A)$ and $B \cup \short(B)$, and between $\short(B)$ and $A \cup
\short(A)$.
\end{packed_itemize}
Fig.~\ref{fig:split_representations} shows two possible representations of a split. Notice that a
split is uniquely determined just by the sets $A$ and $B$, since $\short(A)$ consists of connected
components of $G \setminus (A \cup B)$ attached to $A$, and $\short(B)$ of those attached to $B$. We
refer to this split as the split \emph{between} $A$ and $B$. Alternatively, a split between $A$ and
$B$ is a cut in $G$ between $A$ and $B$ which is a complete bipartite graph.

The standard assumption is that a split is \emph{non-trivial}, meaning that both sides of the split
have at least two vertices: $|A \cup \short(A)| \ge 2$ and $|B \cup \short(B)| \ge 2$. The reason is
that trivial splits are not very interesting: in every graph $G$, the choice $A =
\{u\}$ and $B = N(u)$ for $u \in V(G)$ forms a trivial split. The goal of split decomposition is to
divide a graph into smaller graphs and trivial splits are not helpful.

One of the novelties of this paper is that we study maximal splits.  A split of $G$ between $A$ and
$B$ is \emph{maximal} if there exists no split of $G$ between $A'$ and $B'$ such that $A \subseteq
A'$, $B \subseteq B'$ and $|A|<|A'|$ or $|B|<|B'|$. Both splits between $A$ and $B$ and between $A'$
and $B'$ are allowed to be trivial. Maximal splits satisfy the following property:

\begin{lemma} \label{lem:maximal_split}
A split between $A$ and $B$ is maximal, if and only if there exists no connected component $C$ in
$\short(A)$ such that each vertex of $C$ is either adjacent to all vertices of $A$, or to none of
them, and similarly for $\short(B)$ and $B$.
\end{lemma}

\begin{proof}
Suppose that such a component $C$ in $\short(A)$ exists, and let $C' \subseteq C$ consists of those
vertices which are adjacent to all vertices of $A$. The split between $A$ and $B$ is not maximal
since $A$ and $B \cup C'$ forms a split, for which $C \setminus C' \subseteq \short(B \cup C')$.
Similarly, if such a component $C$ in $\short(B)$ exists, the split between $A$ and $B$ is not
maximal.

On the other hand, suppose that a split between $A$ and $B$ is not maximal, so there exists a split
between $A'$ and $B'$ such that, without loss of generality, $A \subseteq A'$ and $B \subsetneq B'$.
Since every vertex of $B' \setminus B$ is adjacent to all vertices in $A$, we have $B' \setminus B
\subseteq \short(A)$. Choose an arbitrary $c \in B' \setminus B$ and let $C$ be its connected
component of $\short(A)$. As argued, all vertices of $V(C) \cap B'$ are adjacent to all vertices of
$A$. Since $V(C) \cap B' \ne \empty$ and $V(C) \cap A = \emptyset$, the remaining vertices $V(C)
\setminus B' \subseteq \short(B')$. Therefore, they are non-adjacent to all vertices $A$, and $C$
satisfies the properties from the statement of this lemma.
\end{proof}

We always start with a non-trivial split between $A$ and $B$, and modify it using
Lemma~\ref{lem:maximal_split} into a maximal split which may become trivial. But such a trivial
maximal split has a special structure, described below:

\begin{lemma} \label{lem:articulation}
Let $A$ and $B$ form a non-trivial split and let $A'$ and $B'$ form a trivial maximal split such
that $A \subseteq A'$, $B \subseteq B'$, $A' = \{a\}$, and $\short(A') = \emptyset$. Then $a$ is an
articulation in $G$, i.e., $G \setminus a$ is disconnected.
\end{lemma}

\begin{proof}
Since $A \ne \emptyset$, we have $A = \{a\}$. Since the split between $A$ and $B$ is non-trivial, we
have $\short(A) \ne \emptyset$. Therefore, $a$ is an articulation in $G$ which separates $\short(A)$
from $B \cup \short(B)$.
\end{proof}

In the rest of this section, we examine the recursive structure of every possible representation of
$G$ based on maximal splits. In Section~\ref{sec:split_structure}, we analyze the structure of a
representation of a maximal split. In Section~\ref{sec:conditions_split}, we use it to describe the
structure of all circle representations. The described results still apply to trivial maximal
splits, but are not very helpful. Therefore, in Section~\ref{sec:trivial_maximal_splits}, we give a
different description of all representations based on trivial maximal splits.

\subsection{Structure of a Representation of a Maximal Split} \label{sec:split_structure}

Let $\calR$ be a representation of a graph $G$ with a maximal split between $A$ and $B$.  The
representation $\calR$ corresponds to a unique circular word $\tau$.  We consider the circular
subsequence $\gamma$ of $\tau$ induced  by $A \cup B$. The maximal subwords of $\gamma$ consisting
of vertices of $A$ alternate with the maximal subwords of $\gamma$ consisting of vertices of $B$. We
denote all these maximal subwords $\gamma_1,\dots,\gamma_{2k}$ according to their circular order; so
$\gamma = \gamma_1\gamma_2\cdots\gamma_{2k}$. Without loss of generality, we assume that $\gamma_1$
consists of symbols from $A$. We call $\gamma_i$ an \emph{$A$-word} when $i$ is odd, and a
\emph{$B$-word} when $i$ is even. 

We first investigate for each $\gamma_i$ which symbols it contains.

\begin{lemma} \label{lem:group_ordering}
For the subwords $\gamma_1,\dots,\gamma_{k}$ the following holds:
\begin{packed_enum}
\item[(a)] Each $\gamma_i$ contains each symbol at most once.
\item[(b)] The value of $k$ is even and the opposite words $\gamma_i$ and $\gamma_{i+k}$ contain the
same symbols.  
\item[(c)] Let $i \ne j$. If $x \in \gamma_i$ and $y \in \gamma_j$, then $xy \in E(G)$.
\end{packed_enum}
\end{lemma}

\begin{proof}
(a) For every $a \in A$ and $b \in B$, the fact $ab\in E(G)$ implies that $a$ and $b$ alternate in
the circular word $\gamma$. So if some $\gamma_i$ contains both occurrences of, say, $a$,
then $a$ and $b$ would not alternate in $\gamma$.

(b) Let $\gamma_i$ be, say, an $A$-word. We first prove that all the other occurrences of the
symbols from $\gamma_i$ are contained in one word $\gamma_j$; so we get a matching between the
words. Suppose that this is not true and there is $x \in \gamma_i,\gamma_j$ and $y \in
\gamma_i,\gamma_{j'}$ for distinct $i,j$ and $j'$. There is at least one $B$-word $\gamma_\ell$
placed in between $\gamma_j$ and $\gamma_{j'}$ (in the part of the circle not containing
$\gamma_i$). It is not possible for $z \in \gamma_\ell$ to alternate with both $x$ and $y$, which
contradicts $xz,yz \in E(G)$.

Now, let $\gamma_i$ and $\gamma_j$ be two matched $A$-words.  Then every pair of matched $B$-words
must occur on opposite sides of the circle with respect to $\gamma_i$ and $\gamma_j$.  Therefore the
same number of $B$-words occur on both sides of $\gamma_i$ and $\gamma_{j}$, and thus $j = i+k$.

(c) This is implied by (a) and (b) since the occurrences of $x$ and $y$ alternate in $\gamma$.
\end{proof}

Below, we prove that the structure of a maximal split between $A$ and $B$ greatly restricts possible
representation of the vertices of $\short(A) \cup \short(B)$:

\begin{lemma} \label{lem:connected_component}
Let $\tau$ and $\gamma_1,\dots,\gamma_{2k}$ be defined as above. There exists a unique mapping $f :
\short(A) \cup \short(B) \to \{1,\dots,2k\}$ satisfying the following properties: 
\begin{packed_enum}
\item[(a)] For $c \in \short(A) \cup \short(B)$, let $c\tau_cc\tau'_c$ be the subsequence of $\tau$
induced by $A \cup B \cup \{c\}$. Then either $\tau_c$, or $\tau'_c$ is a subword of
$\gamma_{f(c)}$.  For $c \in \short(A)$, the word $\gamma_{f(c)}$ is an $A$-word, while for $c \in
\short(B)$, it is a $B$-word.
\item[(b)] For each connected component $C$ of $\short(A) \cup \short(B)$, the mapping $f|_C$ is
constant, i.e., for all $c,c' \in C$, we have $f(c) = f(c')$, and we denote the image by $f(C)$.
\end{packed_enum}
\end{lemma}

\begin{proof}
Without loss of generality, we assume that $c \in \short(A)$; a symmetric argument works for $c \in
\short(B)$. We first prove the existence and uniqueness of $\gamma_{f(c)}$ when $c$ is adjacent to some
vertex in $a \in A$. Since $c$ alternates with $a$, both $\tau_c$ and $\tau'_c$ are non-empty.  In
(b), we prove that $f(c) = f(c')$ when $cc' \in E(G)$, so by induction the existence and uniqueness
follows for all vertices of $C$.

(a) Since $c$ alternates with $a \in A$, if such $\gamma_{f(c)}$ exists, then it is an $A$-word.
Since $A$-words and $B$-words alternate in $\gamma = \gamma_1\cdots\gamma_{2k}$, we get that
$\tau_c$ is a subword of some $A$-word $\gamma_i$ if and only if it contains no symbol from $B$.
Since at most one of $\tau_c$ and $\tau'_c$ contains no symbol from $B$, it is easy to see that such
$\gamma_i$ is unique. It remains to prove that it always exists.

Let $C$ be the connected component of $\short(A)$ containing $c$.  For contradiction, suppose that
the property (a) fails for $c$. If property (a) fails for $c$, we have $b \in \tau_c$ and $b'
\in \tau'_c$ such that $b,b' \in B$, $b\not=b'$. Since $bc,b'c \notin E(G)$, we also have $bb'
\notin E(G)$; see Fig.~\ref{fig:connected_components}(a). 

\begin{figure}[t!]
\centering
\includegraphics{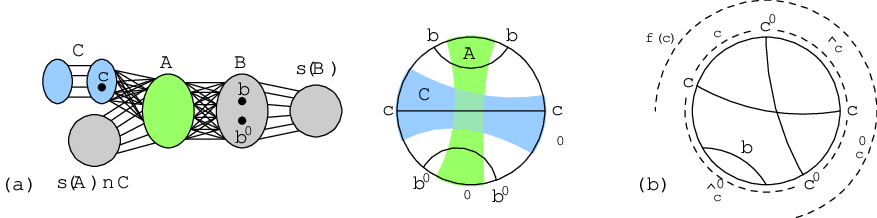}
\caption{(a) On the left, a connected component $C$ of $\short(A)$ attached to $A$. On the right,
the circular subsequences of $\tau$ induced by $A \cup \{b,b',c\}$ and $A \cup \{b,b'\} \cup C$.
(b) The circular subsequence $c\tau_cc'\hat\tau_c c \tau'_cc'\hat\tau'_c$ induced by $A \cup
B \cup \{c,c'\}$, with the word $\gamma_{f(c)}$ depicted. Exactly one of $\tau'_c$ and
$\hat\tau'_c$ contains the symbols $b \in B$.}
\label{fig:connected_components}
\end{figure}
 
For each $a \in A$, we have $ab,ab' \in E(G)$, so $A \cup \{b,b',c\}$ induces in $\tau$ the
subsequence $cb \alpha bcb' \alpha' b'$ such that both $\alpha$ and $\alpha'$ consist only of symbols
from $A$. For each $a \in A$, we have $a \in \alpha$ and $a \in \alpha'$, so $c$ is adjacent to all
vertices of $A$.  Since every vertex $c' \in C$ is connected by a path to $c$, we have that $A \cup
\{b,b'\} \cup C$ induces in $\tau$ the subsequence $\sigma b \alpha b \sigma' b' \alpha' b' $, where
both $\sigma$ and $\sigma'$ consist of symbols of $C$. Therefore every $c' \in C$ is either
adjacent to all vertices of $A$, or to none of them. By, Lemma~\ref{lem:maximal_split}, the split
between $A$ and $B$ is not maximal.

(b) Let $c,c' \in \short(A)$ such that $cc' \in E(C)$ and $f(c)$ is already determined.  We want to
prove that $f(c) = f(c')$. As depicted in Fig.~\ref{fig:connected_components}(b), let
$c\tau_cc'\hat\tau_c c \tau'_cc'\hat\tau'_c$ be the subsequence of $\tau$ induced by $A \cup B \cup
\{c,c'\}$, and suppose that $\tau_c\hat\tau_c$ is a subword of $\gamma_{f(c)}$. Both $\tau'_c$ and
$\hat\tau'_c$ cannot contain symbols from $B$, otherwise $c'$ alternates with $A$ and the argument
in (a) applies. Therefore, either $\hat\tau_c\tau'_c$, or $\hat\tau'_c \tau_c$ is a subword of
$\gamma_{f(c)}$, so $f(c) = f(c')$. We note that either $\hat\tau_c\tau'_c$, or $\hat\tau'_c \tau_c$
might be empty, so $f(c')$ could be chosen arbitrarily to satisfy (a). We then set $f(c') = f(c)$ to
also satisfy (b).
\end{proof}

Let $\tau_i$ denote the subsequence of $\tau$ formed by $\gamma_i$, and of the symbols of
$\bigcup_{C : f(C)=i} V(C)$ over all connected components $C$ of $\short(A) \cup \short(B)$.
By Lemma~\ref{lem:connected_component}, the only difference between $\gamma$ and $\tau$ is that each
subword $\gamma_i$ is replaced by the subword $\tau_i$ which additionally contains all occurrences
of the vertices in some connected components of $\short(A)$ or $\short(B)$. Thus, $\tau =
\tau_1\tau_2\cdots\tau_{2k}$.

Lemma~\ref{lem:connected_component} explains the following naming convention used for maximal splits
between $A$ and $B$ in this paper; see Fig.~\ref{fig:split_representations}. We call the vertices of
$A$ and $B$ as \emph{long vertices} with respect to the maximal split between $A$ and $B$ since each
is represented by ``long chords'' between $\tau_i$ and $\tau_{i+k}$. The vertices $\short(A)$ and
$\short(B)$ are called \emph{short vertices} with respect to the maximal split between $A$ and $B$,
because each is represented by ``short chords'' inside some $\tau_i$.  In the sequel, if the maximal
split is clear from the context, we will just call some vertices long and some vertices short.

\begin{lemma} \label{lem:short_vertex_paths}
If two long vertices $x,y \in A \cup B$ are connected by a path of length at least two having the
internal vertices in $\short(A) \cup \short(B)$, then $x$ and $y$ belong to the same pair $\gamma_i$
and $\gamma_{i+k}$ in every representation.
\end{lemma}

\begin{proof}
Let $C$ be the connected component of $\short(A) \cup \short(B)$ having the internal vertices of
this path between $x$ and $y$. By Lemma~\ref{lem:connected_component}, all vertices of $C$ have both
symbols in $\tau_{f(C)}$.  Therefore, $x,y \in \tau_{f(C)}$. So $x,y \in \gamma_{f(C)}$, and by
Lemma~\ref{lem:group_ordering}(b) also $x,y \in \gamma_{f(C)+k}$.
\end{proof}

Also, we prove the following simple lemma:

\begin{lemma} \label{lem:fourclique}
Let $x$, $y$, $z$, and $w$ be distinct vertices inducing a clique in $G$, and let $P$ be a path from
$x$ to $y$ of length at least 2. If $xzywxzyw$ is a subsequence of the circular word $\tau$ of a
circle representation of $G$, then some internal vertex of $P$ is adjacent to $z$ or $w$.
\end{lemma}

\begin{proof}
Let $v_1,\dots,v_k$ be the internal vertices of $P$ such that $v_1x \in E(G)$. We prove by induction
that no $v_i$ having $v_iz \in E(G)$ or $v_iw \in E(G)$ implies that $v_ky \notin E(G)$. If $v_1$ is
not such a vertex, then we get that $v_1xv_1zywxzyw$ is a subsequence of $\tau$ since $v_1x \in
E(G)$. For the induction hypothesis, suppose that $v_iv_izywzyw$ is a subsequence of $\tau$. Since
$v_iv_{i+1} \in E(G)$, if $v_{i+1}$ is not adjacent to $z$ and $w$, we get that
$v_{i+1}v_{i+1}zywzyw$ is a subsequence of $\tau$.  Therefore, $v_k$ does not alternate with $y$,
contradicting that $v_ky \in E(G)$.
\end{proof}

\subsection{Conditions Forced by a Maximal Split} \label{sec:conditions_split}

Now, we want to investigate the opposite relation. Namely, what can one say about a representation
from the structure of a maximal split? Suppose that $x$ and $y$ are two long vertices.  We want to
know the properties of $x$ and $y$ which force every representation $\calR$ to have a subword
$\gamma_i$ of $\gamma$ containing both $x$ and $y$.

Inspired by Naji~\cite[Section IV.4]{N85}, we define a symmetric relation $\sim$ on $A \cup B$ where
$x \sim y$ means that $x$ and $y$ must occur in the same subword $\gamma_i$ of $\gamma$.  This
relation is given by two conditions:
\begin{packed_head_enum}{(C1)}
\item[(C1)] Lemma~\ref{lem:group_ordering}(c) states that if $xy \notin E(G)$, then $x
\sim y$, i.e., if $x$ and $y$ are placed in different subwords, then $C_x$ intersects $C_y$.
In particular, $x\sim x$.
\item[(C2)] Lemma~\ref{lem:short_vertex_paths} gives $x \sim y$ when $x$ and $y$ are
connected by a non-trivial path with all the inner vertices in $\short(A) \cup \short(B)$.
\end{packed_head_enum}
Let us take the transitive closure of $\sim$, which we denote by $\sim$ thereby slightly abusing the
notation.  Thus, we obtain an equivalence relation $\sim$ on $A\cup B$.  Notice that every
equivalence class of $\sim$ is either fully contained in $A$ or in $B$.
Figure~\ref{fig:split_representations} on right shows schematically a situation in which the
relation $\sim$ has four equivalence classes $A_1$, $A_2$, $B_1$ and $B_2$.

Now, let $\Phi$ be an equivalence class of $\sim$. We denote by $\short(\Phi)$ the set consisting of
all the vertices in the connected components of $\short(A) \cup \short(B)$ which have a vertex
adjacent to a vertex of $\Phi$. Since $\sim$ satisfies (C2), we know that the sets $\short(\Phi)$ of
the equivalence classes of $\sim$ define a partition of $\short(A) \cup \short(B)$.

\heading{Recognition Algorithms Based on Splits.}
Split decompositions are used in the current state-of-the-art algorithms for recognizing circle
graphs.  If a circle graph contains no split, it is called a \emph{prime graph}. The representation
of a prime graph is uniquely determined (up to the orientation of the circle) and can be constructed
efficiently. There is an algorithm  which finds a split in a graph in linear time~\cite{dalhaus}. In
fact, the entire \emph{split decomposition tree} (i.e., the recursive decomposition tree obtained
via splits) can be found in linear time. Usually the representation $\calR$ is constructed as
follows.

\begin{figure}[t!]
\centering
\includegraphics{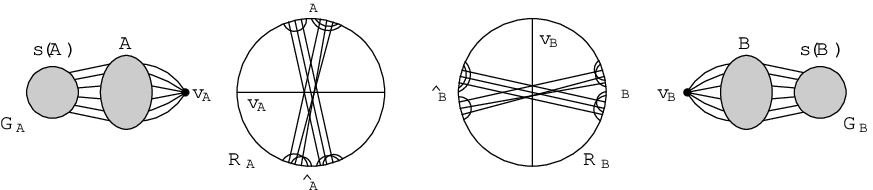}
\caption{The graphs $G_A$ and $G_B$ together with some constructed representations $\calR_A$ and
$\calR_B$. By joining these representations, we get the representation shown on the left in
Fig.~\ref{fig:split_representations}.}
\label{fig:recursion_on_split}
\end{figure}

We define two graphs $G_A$ and $G_B$ where $G_A$ is created from $G$ by contracting the vertices of
$B \cup \short(B)$ into a new vertex $v_A$ and $G_B$ by contracting $A \cup \short(A)$ into a new
vertex $v_B$. So $v_A$ is adjacent to all vertices in $A$ and to no vertices in $\short(A)$, and
similarly for $v_B$. Then we apply the algorithm recursively on $G_A$ and $G_B$ and construct their
representations $\calR_A$ and $\calR_B$; see Fig.~\ref{fig:recursion_on_split}.  It remains to join
the representations $\calR_A$ and $\calR_B$ in order to construct $\calR$.

To this end we take $\calR_A$ and replace $C_{v_A}$ by the representation of $B \cup \short(B)$ in
$\calR_B$. More precisely, let the circular ordering of the endpoints of chords defined by $\calR_A$
be $v_A \tau_A v_A \hat\tau_A$ and let the circular ordering defined by $\calR_B$ be $v_B \tau_B v_B
\hat\tau_B$. The constructed $\calR$ has the corresponding circular ordering
$\tau_A\tau_B\hat\tau_A\hat\tau_B$. It is easy to see  that $\calR$ is a correct circle
representation of $G$.

\heading{Structure of All Representations.}
The above algorithm constructs a very specific representation $\calR$ of $G$, and a representation
like the one in Fig.~\ref{fig:split_representations} on the right cannot be constructed in this way
using the split between $A$ and $B$. In what follows we describe the structure of all the
representations of a circle graph $G$ based on the different circular orderings of the equivalence
classes of $\sim$. While the described structure of all the representations depends on the maximal
split that we chose, the relation $\sim$ defined with respect to this maximal split can be used to
generate all the representations of $G$.

We choose an arbitrary circular ordering $\Phi_1,\dots,\Phi_\ell$ of the classes of $\sim$.  Let
$G_i$ be a graph constructed from $G$ by contracting the vertices $V(G) \setminus \bigl(\Phi_i \cup
\short(\Phi_i)\bigr)$ into one vertex $v_i$; i.e., $G_i$ is defined similarly to $G_A$ and $G_B$
above.  Let $\calR_1,\dots,\calR_\ell$ be arbitrary representations of $G_1,\dots,G_\ell$. We join
these representations as follows. Let $v_i \tau_i v_i \hat\tau_i$ be the circular ordering of
$\calR_i$.  We construct $\calR$ as the circular ordering
\begin{equation} \label{eq:representation}
\tau_1\tau_2\dots \tau_{\ell-1}\tau_\ell\hat\tau_1\hat\tau_2 \dots \hat\tau_{\ell-1}\hat\tau_\ell.
\end{equation}
In Fig.~\ref{fig:split_representations}, we obtain the representation on the left by the circular
ordering $A_1 A_2 B_1 B_2$ of the classes of $\sim$ and the representation on the right by $A_1 B_1
A_2 B_2$.

First, we show that every representation obtained in this way is correct.

\begin{lemma} \label{lem:rep_correctness}
Every circular ordering (\ref{eq:representation}) constructed as above defines a circle
representation of $G$.
\end{lemma}

\begin{proof}
Let $u,v \in V(G)$. We shall prove that $u$ and $v$ are adjacent in $G$ if and only if they
alternate in $\calR$. Suppose that $u,v \in V(G_i) \setminus \{v_i\}$. Since the cyclic subsequence
$\tau_i\hat{\tau}_i$ appears in both $\calR_i$ and $\calR$, two vertices in $V(G_i)\setminus
\{v_i\}$ alternate in $\calR$ if and only if they are adjacent in $G_i$, which is if and only if
they are adjacent in $G$.

Otherwise, let $u \in V(G_i) \setminus \{v_i\}$ and $v \in V(G_j)\setminus \{v_j\}$ for $i \ne j$.
Then $uv \in E(G)$ if and only if they are both long vertices. Each long vertex of $\Phi_t$ appears
once in both $\tau_t$ and $\hat{\tau}_t$, but each short vertex $\short(\Phi_t)$ has both its
occurrences either in $\tau_t$, or in $\hat{\tau}_t$. We conclude that $u$ and $v$ alternate in $\calR$
if and only if they are both long vertices, i.e., if and only if they are adjacent in $G$ since $u$
and $v$ do not satisfy (C1).
\end{proof}

Next, we analyze every representation $\calR$ of $G$.

\begin{lemma} \label{lem:consecutive_classes}
Let $\tau$ be the circular word corresponding to a representation $\calR$ of $G$.  Then the symbols
of $\Phi_i \cup \short(\Phi_i)$ form exactly two subwords $\tau_i$ and $\hat\tau_i$ of $\tau$ such
that for each $u \in \Phi_i$, we have $u \in \tau_i$ and $u \in \hat\tau_i$, while each $v \in
\short(\Phi_i)$ has both endpoint either in $\tau_i$, or in $\hat\tau_i$.
\end{lemma}

\begin{proof}
Let $\calR$ be a representation of $G$ and consider how it represents $A \cup B$. We get the
subwords $\gamma_1,\dots,\gamma_{2k}$ of the endpoints of $A \cup B$, as described in
Section~\ref{sec:split_structure}.

Let $x \in \Phi_i$ such that $x \in \gamma_j$. We claim that $\Phi_i$ is a subset of $\gamma_j$.
Since $\Phi_i$ is an equivalence class of~$\sim$, let $y \in \Phi_i$ such that one of the conditions
(C1) or (C2) applies to $x$ and $y$. Since $\sim$ is the transitive closure of conditions (C1) and
(C2), to prove the claim, it is sufficient to show that $y \in \gamma_j$. If (C1) applies, then
$y\in\gamma_j$ by Lemma~\ref{lem:group_ordering}(c).  If (C2) applies, then $y \in \gamma_j$ by
Lemma~\ref{lem:short_vertex_paths}. By Lemma~\ref{lem:group_ordering}(a), each vertex of
$\Phi_i$ appears exactly once in $\gamma_j$ and once in $\gamma_{j+k}$.

Furthermore, we claim that the vertices of $\Phi_i$ form subwords of $\gamma_j$ and $\gamma_{j+k}$.
Let $z \in \gamma_j$ be placed between $x\in \Phi_i$ and $y\in \Phi_i$.  First, we assume that (C1)
or (C2) applies to $x$ and $y$.
\begin{packed_itemize}
\item If (C1) applies to $x$ and $y$, then $xy \notin E(G)$. As $x$ and $y$ do
not alternate, it is not possible for $z$ to alternate with both $x$ and $y$. Thus $z \sim x$ or $z
\sim y$, which in turn implies that $z \in \Phi_i$.
\item Suppose that (C2) applies to $x$ and $y$. If $xz \notin E(G)$ or $yz \notin E(G)$, we get that
$z \in \Phi_i$ by (C1). Otherwise, we claim that a path $P$ from $x$ to $y$ having all the internal
vertices in $\short(\Phi_i)$ has at least one internal vertex adjacent to $z$. For every $w \in
\gamma_{j+1}$ and $w \in \gamma_{j+1+k}$, we have $xw,yw,zw \in E(G)$, but none of the inner
vertices of $P$ are adjacent to $w$.  Since $\{x,y,z,w\}$ induce the subsequence $xzywxzyw$ in
$\tau$, by Lemma~\ref{lem:fourclique} some inner vertex $P$ has to alternate with $z$. Thus, $z\sim
x$ and $z\sim y$ by (C2), so $z \in \Phi_i$.
\end{packed_itemize}
If $x\sim y$ and neither of (C1) and (C2) applies, we easily proceed by an inductive argument on the
number of applications of (C1) and (C2).  If $x\sim y'\sim y$ and a vertex $z \in \gamma_j$ is
placed between $x$ and $y$ in $\gamma_j$, then $z$ is also placed in $\gamma_j$ between $x$ and $y'$
or between $y'$ and $y$.

By the above argument, each class $\Phi_i$ forms two subwords of $\gamma$. By adding the
short vertices $\short(\Phi_i)$ as in Lemma~\ref{lem:connected_component} applied on the maximal
split between $\Phi_i$ and $A \cup B \setminus \Phi_i$, we obtain two subwords of $\tau$ for each
class $\Phi_i$.
\end{proof}

Now, we are ready to prove the main structural proposition.

\begin{proposition} \label{prop:structure}
Let $A$ and $B$ form a maximal split of $G$ and let $\sim$ be the equivalence relation defined by
(C1) and (C2) on $A \cup B$. Then every representation $\calR$ of $G$ corresponds to some circular
ordering $\Phi_1,\dots,\Phi_\ell$ and to some representations $\calR_1,\dots,\calR_\ell$ of
$G_1,\dots,G_\ell$. More precisely, $\calR$ can be constructed by arranging
$\calR_1,\dots,\calR_\ell$ as in (\ref{eq:representation}):
$\tau_1\dots\tau_\ell\hat\tau_1\dots\hat\tau_\ell$.
\end{proposition}

\begin{proof}
By Lemma~\ref{lem:rep_correctness}, every representation constructed by~(\ref{eq:representation}) is
correct. On the other hand, let $\calR$ be a representation of $G$ with the corresponding circular
word $\tau$. According to Lemma~\ref{lem:consecutive_classes}, we know that $\Phi_i \cup \short(\Phi_i)$
forms two subwords $\tau_i$ and $\hat\tau_i$ of $\tau$.  For $i \ne j$, the edges between $\Phi_i$
and $\Phi_j$ form a complete bipartite graph.  The subwords $\tau_i$, $\hat\tau_i$, $\tau_j$ and
$\hat\tau_j$ alternate, i.e., appear as $\tau_i\tau_j\hat\tau_i\hat\tau_j$ or
$\tau_j\tau_i\hat\tau_j\hat\tau_i$ in $\tau$.  Thus, if we start from some point along the circle,
the order of $\tau_i$'s gives a circular ordering $\Phi_1,\dots,\Phi_\ell$ of the classes. The
representation $\calR_i$ has the circular word $v_i\tau_iv_i\hat\tau_i$.
\end{proof}

\subsection{The Structure of All Representations of Trivial Maximal Splits}
\label{sec:trivial_maximal_splits}

Let $A$ and $B$ form a trivial maximal split with $A = \{a\}$ and $\short(A) = \emptyset$, created
from a non-trivial split. The results described in Sections~\ref{sec:split_structure}
and~\ref{sec:conditions_split} still apply to this split, but they are not very helpful. By
Lemma~\ref{lem:articulation}, $a$ is an articulation in $G$. So, $G[B]$ consists of at least
two connected components and $\sim$ has two equivalence classes $\Phi_1 = A$ and $\Phi_2 = B$. Since
$G_2 \cong G$, Proposition~\ref{prop:structure} describes all representations of $G$ in terms of all
representations of $G$.

In this section, we show that all possible representations can be easily described in a different
way, based on all different representations of connected components of $G \setminus a$.  The
following lemma states that connected components do not alternate in any circle representation:

\begin{lemma} \label{lem:disconnected}
Let $C$ and $C'$ be two distinct connected component of a circle graph $G$. No representation
has a subword $uxvy$ where $u,v \in V(C)$ and $x,y \in V(C')$.
\end{lemma}

\begin{proof}
Let $\sigma$ be the subsequence induced by $V(C) \cup V(C')$. We know that $\sigma = \sigma_1 \cdots
\sigma_{2k}$ such that $\sigma_i$ is the maximal subword consisting only of symbols from $V(C)$ if
$i$ is odd, and only of symbols from $V(C')$ if $i$ is even. We want to prove that $k=1$. For
contradiction, suppose that $k > 1$. Since $C$ is connected, there exists $u \in C$ such that $u \in
\sigma_1$ and $u \in \sigma_i$ for $i > 1$. Since $C'$ is connected, there exists $x \in C'$ such
that $x \in \sigma_2 \sigma_4 \cdots \sigma_{i-1}$ and $x \in \sigma_{i+1} \cdots \sigma_{2k}$.
Since $u$ and $x$ alternate, we have $ux \in E(G)$ which is a contradiction.
\end{proof}

We choose an arbitrary ordering of the connected components of $G \setminus a$ as $C_1,\dots,C_\ell$.
Let $G_i$ be the subgraph of $G$ induced by $\{a\} \cup V(C_i)$. Let $\calR_i$ be an arbitrary
representation of $G_i$ having the circular word $a\tau_ia\hat\tau_i$. We construct the joined
representation $\calR$ of $G$ by the circular word
\begin{equation} \label{eq:art_representation}
a\tau_1\tau_2\cdots\tau_{\ell-1}\tau_\ell a\hat\tau_\ell\hat\tau_{\ell-1}\cdots\hat\tau_2\hat\tau_1;
\end{equation}
see Fig.~\ref{fig:trivial_split_representations}. First, we prove that every such constructed
representation of $G$ is correct:

\begin{figure}[t!]
\centering
\includegraphics{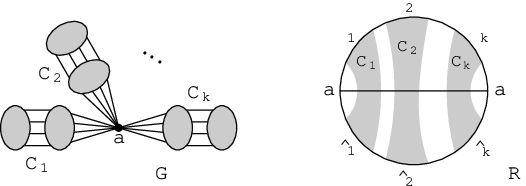}
\caption{If $a$ is an articulation, then every circle representation $\calR$ of $G$ consists of some
ordering of connected components $C_1,\dots,C_\ell$ of $G \setminus a$ and it corresponds to the
depicted circular word $\tau$ in which $a\tau_ia\hat\tau_i$ is some representation of the subgraph
of $G$ induced by $V(C_i) \cup \{a\}$.}
\label{fig:trivial_split_representations}
\end{figure}

\begin{lemma} \label{lem:art_rep_correctness}
Every circular ordering (\ref{eq:art_representation}) constructed as above defines a circle
representation of $G$.
\end{lemma}

\begin{proof}
Let $\tau$ be the circular ordering constructed using~(\ref{eq:art_representation}). Since $V(G_i)$
induces the subsequence $a \tau_i a \hat\tau_i$, each $G_i$ is represented correctly in $\calR_i$.
For $i < j$, the vertices of $V(C_i) \cup V(C_j)$ induce in $\tau$ the subsequence
$\tau_i\tau_j\hat\tau_j\hat\tau_i$, so no two vertices $u \in V(C_i)$ and $v \in V(C_j)$ alternate
and the non-edges between $C_i$ and $C_j$ are represented correctly. 
\end{proof}

Next, we analyze every representation $\calR$ of $G$.

\begin{lemma} \label{lem:art_consecutive_classes}
Let $\tau$ be the circular word corresponding to a representation $\calR$ of $G$.  Then the symbols
of $V(C_i)$ form exactly two subwords $\tau_i$ and $\hat\tau_i$ of $\tau$ such that
$a\tau_ia\hat\tau_i$ is a subsequence of $\tau$.
\end{lemma}

\begin{proof}
Since $V(C_i) \cap B \ne \emptyset$, there exists some $b \in V(C_i)$ which alternates with $a$, so
the symbols of $V(C_i)$ form at least two subwords alternating with $a$. If $V(C_i)$ would form more
than two subwords, then $\tau$ has a subsequence $auxva$, where $u,v \in V(C_i)$ and $x \in V(C_j)$
for $j \ne i$. Since some $y \in V(C_j)$ alternates with $a$, it follows that $\tau$ has the
subsequence $auxvay$, so we get $uxvy$ which is not possible by Lemma~\ref{lem:disconnected}.
\end{proof}

Now, we are ready to prove the following structural proposition.

\begin{proposition} \label{prop:art_structure}
Let $A = \{a\}$ and $B$ form a trivial maximal split of $G$ created from a non-trivial split.  Then
every representation $\calR$ of $G$ corresponds to some ordering $C_1,\dots,C_\ell$ of connected
components of $G \setminus a$ and to some representations $\calR_1,\dots,\calR_\ell$ of
$G_1,\dots,G_\ell$ where $G_i$ is the subgraph of $G$ induced by $V(C_i) \cup \{a\}$. More
precisely, $\calR$ can be constructed by arranging $\calR_1,\dots,\calR_\ell$ as in
(\ref{eq:art_representation}): $a \tau_1\dots\tau_\ell a \hat\tau_\ell\dots\hat\tau_1$.
\end{proposition}

\begin{proof}
By Lemma~\ref{lem:art_rep_correctness}, every representation constructed
by~(\ref{eq:art_representation}) is correct. On the other hand, let $\calR$ be a representation of
$G$ corresponding to a circular word $\tau$. Suppose that $G \setminus a$ has $\ell$ connected
components. The circular word $\tau$ defines an ordering $C_1,\dots,C_\ell$ of the connected components
of $G \setminus a$ in the following way. By Lemma~\ref{lem:art_consecutive_classes}, $\tau = a
\tau_1 \tau_2 \cdots \tau_\ell a \sigma_\ell \sigma_{\ell-1} \cdots \sigma_1$, where $\tau_i$ and precisely
one $\sigma_j$ are two maximal subwords of $\tau$ containing all symbols from $V(C_i)$.  Since the
connected components $C_1,\dots,C_\ell$ cannot alternate by Lemma~\ref{lem:disconnected}, we get that
$\sigma_i$ consists of symbols of $C_i$, i.e., $\sigma_i = \hat\tau_i$. Each $a\tau_ia\hat\tau_i$
gives some representation $\calR_i$ of $G_i$.
\end{proof}

\section{Algorithm} \label{sec:algorithm}

In this section, we describe an $\O(n^3)$ algorithm for the partial representation extension problem
of circle graphs. It is based on the structure of all representations of
Section~\ref{sec:structural_results}. Recall that a partial representation $\calR'$ gives a circular
word $\tau'$ such that each vertex $u \in V(G)$ appears at most twice in $\tau'$. We want to decide
whether there exists a representation $\calR$ corresponding to a circular word $\tau$ such that
$\tau'$ is a subsequence of $\tau$.

\heading{Dealing with Disconnected Graphs.}
To apply the structural properties of Section~\ref{sec:structural_results}, we need to work with
connected graphs. In general, the partial representation extension problems cannot be trivially
restricted to connected inputs, as in the case of most graph problems.  In particular, for some
classes the problems are polynomial-time solvable for connected inputs and \cFPT\ in the number of
components for disconnected inputs, but \cNP-complete in general; see e.g. ~\cite{KKORSSV,KKOS}.
The reason is that the components are placed together in one representation and they restrict each
other.

In the case of circle graphs, we can deal with disconnected inputs easily. By
Lemma~\ref{lem:disconnected}, we know that $\tau'$ cannot contain a subsequence $uxvy$ where $u,v$
belong to one component and $x,y$ to another one. If this happens, we immediately output ``no''.
Otherwise the question of extendibility is equivalent to testing whether each component $C$ is
extendible where the partial representation of $C$ is given by the subsequence of $\tau'$ containing
all occurrences of the vertices of $C$. So from now on we assume that the input graph $G$ is
connected.

\heading{Overview.}
Our algorithm proceeds recursively via split decomposition. For each encountered graph $G$ with a
partial representation $\calR'$ corresponding to the circular word $\tau'$, it proceeds with the
following steps: 
\begin{packed_itemize}
\item If $G$ is prime, we have two possible representations (one is reversal of the other) and we
test whether one of them extends $\tau'$. We return the result.
\item Otherwise, we find a non-trivial split and modify it into a maximal split between $A$ and
$B$, using Lemma~\ref{lem:maximal_split}. Next, we proceed with one of the following steps.
\item In Case I, the maximal split between $A$ and $B$ is non-trivial. We compute the relation
$\sim$.  We try to determine an ordering $\Phi_1,\dots,\Phi_\ell$ of the equivalence classes of
$\sim$ along the circle as in~(\ref{eq:representation}) which is compatible with the partial
representation $\calR'$. This order is partially prescribed by pre-drawn endpoints of short and long
vertices and we recurse on testing whether partial representations of different equivalence classes
$\Phi \cup \short(\Phi)$ can be extended. If no ordering is compatible, we stop and output ``no''.
\item In Case II, the maximal split between $A$ and $B$ is trivial with $A = \{a\}$ and $\short(A) =
\emptyset$. We try to determine an ordering $C_1,\dots,C_k$ of the connected components of $G
\setminus a$ along the circle as in~(\ref{eq:art_representation}) which is compatible with the
partial representation $\calR'$. This order is partially prescribed by pre-drawn endpoints of chords
and we recurse on testing whether partial representations of different components $C$ can be
extended. If no ordering is compatible, we stop and output ``no''.
\end{packed_itemize}
For a more detailed overview of the main steps, see Algorithm~\ref{alg:overview}. Now we describe
everything in detail.

\begin{algorithm}[t!]
\centering
\begin{tabbing}
\textbf{Input:}\enspace A circle graph $G$ and a partial representation $\calR'$ corresponding to a
circular word $\tau'$.\\
\textbf{Output:}\enspace \ACCEPT if $\calR'$ is extendible, \REJECT otherwise.\\[1.5ex]
1.\quad \=\textbf{If} \= $\calR'$ is incorrect \textbf{then} \REJECT.\\
2. \>\textbf{If} $G$ is a prime graph \textbf{then}\\
3. \>\> Construct the unique representations $\tau$ and (its reverse) $\tau_R$ of $G$.\\
4. \>\> \textbf{If} \= $\tau'$ is a subsequence of $\tau$ or $\tau_R$
	\textbf{then} \ACCEPT \textbf{else} \REJECT. \\
5. \>\textbf{Else} ($G$ is not a prime graph)\\
6. \>\> Find a non-trivial split between $A'$ and $B'$.\\
7. \>\> Modify it into a maximal split between $A$ and $B$ such that $A' \subseteq A$ and $B'
\subseteq B$.\\[1ex]
8. \>\> \textbf{Case I: If} the maximal split between $A$ and $B$ is non-trivial \textbf{then}\\
9. \>\>\vrule\> Compute the equivalence relation $\sim$.\\
10. \>\>\vrule\> Let $\tau' = \tau'_1\cdots\tau'_k$ be the maximal subwords of extended classes $\Psi$.\\
11. \>\>\vrule\> \textbf{Ca}\=\textbf{se I.1: If} some extended class corresponds to two maximal
subwords in $\tau'$ \textbf{then}\\
12. \>\>\vrule\>\> Compute a circular ordering $\Psi_1,\dots,\Psi_\ell$ compatible with $\tau'$.\\
13. \>\>\vrule\>\> Construct the partial representations $\calR'_i$ of $G_i$.\\
14. \>\>\vrule\>\> \textbf{If} all $\calR'_1,\dots,\calR'_\ell$ are extendible 
	\textbf{then} \ACCEPT \textbf{else} \REJECT. \\
15. \>\>\vrule\> \textbf{Case I.2: Else} (each extended class corresponds to one maximal subword in $\tau'$) \\
16. \>\>\vrule\>\> Construct the partial representations $\calR'_i$ and $\widetilde\calR'_i$ of $G_i$.\\
17. \>\>\vrule\>\> Proceed as in the subroutine of Algorithm~\ref{alg:caseI2}.\\[1ex]
18. \>\> \textbf{Case II: Else} (the maximal split between $A$ and $B$ is trivial
	with $A = \{a\}$ and $\short(A) = \emptyset$) \\
19. \>\>\vrule\> Compute the connected components of $G \setminus a$.\\
20. \>\>\vrule\> \textbf{Case II.1: If} both endpoints of $a$ appear in $\tau'$ \textbf{then} \\
21. \>\>\vrule\>\> Compute a linear ordering $C_1,\dots,C_\ell$ compatible with $\tau'$.\\
22. \>\>\vrule\>\> Construct the partial representations $\calR'_i$ of $G_i$.\\
23. \>\>\vrule\>\> \textbf{If} all $\calR'_1,\dots,\calR'_\ell$ are extendible 
	\textbf{then} \ACCEPT \textbf{else} \REJECT. \\
24. \>\>\vrule\> \textbf{Case II.2: Else if} single endpoint of $a$ appears in $\tau'$ \textbf{then} \\
25. \>\>\vrule\>\> Decompose the problem into two subproblems.\\
26. \>\>\vrule\>\> One is solved using Case II.1, the other as in Case I.2.\\
27. \>\>\vrule\>\> \textbf{If} both succeed	\textbf{then} \ACCEPT \textbf{else} \REJECT. \\
28. \>\>\vrule\> \textbf{Case II.3: Else} (no endpoint of $a$ appears in $\tau'$) \\
29. \>\>\vrule\>\> \textbf{Ca}\=\textbf{se II.3a: If} some component has two maximal subwords in $\tau'$
\textbf{then}\\ 
30. \>\>\vrule\>\>\> Decompose the problem into three subproblems.\\
31. \>\>\vrule\>\>\> Two are solved using Case II.2, the last one using Case II.1.\\
32. \>\>\vrule\>\>\> \textbf{If} all succeed \textbf{then} \ACCEPT \textbf{else} \REJECT. \\
33. \>\>\vrule\>\> \textbf{Else} (no component has two maximal subwords in $\tau'$)\\
34. \>\>\vrule\>\>\> Proceed as in the subroutine of Algorithm~\ref{alg:caseII3}.\\[-0.5ex]
\end{tabbing}
\vskip -1.2em
\caption{The $\O(n^3)$ algorithm for $\ext(\circle)$.}
\label{alg:overview}
\end{algorithm}

\heading{Testing Correctness of $\boldsymbol{\calR'}$.} In the beginning, the algorithm tests
correctness of the input partial representation. If $u,v \in V(G)$ have both occurrences in $\tau'$,
we check that these occurrences alternate if and only if $uv \in E(G)$, and if some pair is
represented incorrectly, we stop the algorithm and output ``no''. If only a single endpoint of $u
\in V(G)$ appears in $\tau'$, no checking is done. This checking can be done trivially in time
$\O(n^2)$.

\heading{Prime Graphs.}
A graph is called \emph{prime} if it contains no split.  If $G$ is a prime graph, then it has at
most two different representations $\calR$ and $\hat\calR$~\cite{dalhaus} where one is the reversal
of the other.  We just need to test whether one of them extends $\calR'$.  We can construct one of
these representations in quasilinear time~\cite{GPTC13}.

\heading{Finding a Maximal Split Between $\boldsymbol A$ and $\boldsymbol B$.}
If the graph $G$ is not prime, then we can find a non-trivial split between $A'$ and $B'$ in linear
time~\cite{dalhaus}. Using Lemma~\ref{lem:maximal_split}, we modify it into a maximal split between
$A$ and $B$ such that $A' \subseteq A$ and $B' \subseteq B$ in linear time.

\subsection{Case I: A Non-trivial Maximal Split Between $\boldsymbol A$ and $\boldsymbol B$.}
\label{sec:caseI}

We start by computing the equivalence relation $\sim$ which can be done in time $\O(n^2)$. Next, we
want to find an ordering of its equivalence classes. For a class $\Phi$ of $\sim$, we define the
\emph{extended class} $\Psi$ of $\sim$ as $\Phi \cup \short(\Phi)$.  If some extended class has no
vertex pre-drawn, we may choose an arbitrary representation and place it in an arbitrary order,
so we can ignore such classes for the rest of Case I. Let $\sim$ have $\ell$ equivalence classes,
all of them appearing in $\tau'$.

The circular word $\tau'$ is composed of $k$ \emph{maximal subwords} $\tau' = \tau'_1 \tau'_2 \cdots
\tau'_k$ such that each $\tau'_i$ contains only symbols of one extended class $\Psi$. According to
Proposition~\ref{prop:structure}, each extended class $\Psi$ corresponds to at most two different
maximal subwords.  Also, if two extended classes $\Psi$ and $\hat\Psi$ each correspond to two
different maximal subwords, then occurrences of these subwords alternate in $\tau'$. Otherwise we
reject the input.

\begin{figure}[t!]
\centering
\includegraphics{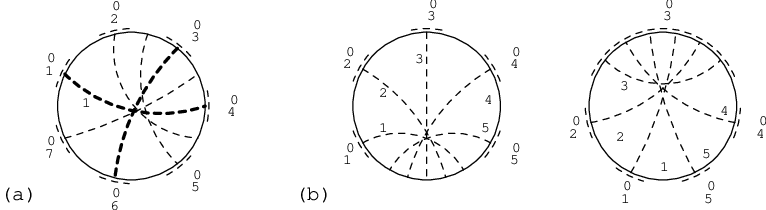}
\caption[]{Each dashed line represents one extended class. (a) An example of Case I.1. We have
$\tau' = \tau'_1\cdots\tau'_7$ and five extended classes $\Psi_1$ (corresponding to $\tau'_1$ and
$\tau'_4$), $\Psi_\alpha$ (to $\tau'_2$), $\Psi_\beta$ (to $\tau'_3$ and $\tau'_6$), $\Psi_\gamma$
(to $\tau'_5$), and $\Psi_\delta$ (to $\tau'_7$). We get that $\Psi_1 < \Psi_\alpha < \Psi_\beta$
and $\Psi_1 < \Psi_\gamma < \Psi_\beta < \Psi_\delta$, so one possible circular ordering is
$\Psi_1,\Psi_\alpha,\Psi_\gamma,\Psi_\beta,\Psi_\delta$, and $\alpha=2$, $\gamma=3$, $\beta=4$, and
$\delta=5$. By Lemma~\ref{lem:correctness_of_caseI1}, $\calR'$ is extendible if and only if
$\calR'_1,\dots,\calR'_5$ are extendible.
(b) Examples of two possible extending representations in Case I.2. On the left, $\tau'_i$ is
extended by $\tau_i$ in an extending representation $\calR$, which is possible only when
$\calR'_1,\dots,\calR'_5$ are extendible. On the right, $\tau'_3$ is extended by both $\tau_3$ and
$\hat\tau_3$. By Lemma~\ref{lem:two_rep_cases}, $\calR'_1,\calR'_2,\calR'_4,\calR'_5$ are
extendible, but it is sufficient for $\widetilde\calR'_3$ to be extendible.}
\label{fig:caseI}
\end{figure}

\heading{Case I.1: An extended class corresponds to two maximal subwords.}\\
We denote this class by $\Psi_1$ and put this class as first in the ordering. By renumbering, we may
assume that $\Psi_1$ corresponds to $\tau'_1$ and $\tau'_t$. Then one circular order of the classes
can be determined by the following linear ordering $<$ starting with $\Psi_1$. Let $\Psi_i$ and
$\Psi_j$ be two distinct classes. If $\Psi_i$ corresponds to $\tau'_a$ and $\Psi_j$ corresponds to
$\tau'_b$ such that either $a < b < t$ or $t < a < b$, we put $\Psi_i < \Psi_j$. We obtain the
ordering of the classes as any linear extension of $<$. Since subwords of all extended classes with
two subwords in $\tau'$ alternate, we get that $<$ is acyclic and a linear extension always exists.
Figure~\ref{fig:caseI}(a) shows an example.

We have ordered the extended classes $\Psi_1,\dots,\Psi_\ell$ and the corresponding classes
$\Phi_1,\dots,\Phi_\ell$. We construct each $G_i$ with the vertices $\Psi_i \cup \{v_i\}$ as in
Section~\ref{sec:conditions_split}, so $v_i$ is adjacent to $\Phi_i$ and non-adjacent to
$\short(\Phi_i)$. The partial representation $\calR'_i$ of $G_i$ is either the word
$v_i\tau'_sv_i$ (if $\Psi_i$ corresponds to the single maximal subword $\tau'_s$ in $\tau'$) or 
the word $v_i\tau'_s v_i \tau'_t$ (if $\Psi_i$ corresponds to two maximal subwords $\tau_s'$ and
$\tau_t'$ in $\tau'$). We test recursively, whether each representation $\calR'_i$ of $G_i$ is
extendible to a representation of $\calR_i$. If yes, we join $\calR_1,\dots,\calR_\ell$ as in
Proposition~\ref{prop:structure}. Otherwise, the algorithm outputs ``no''.

\begin{lemma} \label{lem:correctness_of_caseI1}
In Case I.1, the representation $\calR'$ is extendible if and only if the representations
$\calR'_1,\dots,\calR'_\ell$ of the graphs $G_1,\dots,G_\ell$ are extendible.
\end{lemma}

\begin{proof}
Suppose that $\calR$ extends $\calR'$. According to Proposition~\ref{prop:structure}, the
representations of $\Psi_1,\dots,\Psi_\ell$ are ordered along the circle, and so we obtain
representations $\calR_1,\dots,\calR_\ell$ extending $\calR'_1,\dots,\calR'_\ell$.

For the other implication, we just take $\calR_1,\dots,\calR_\ell$ and combine them to form $\calR$ 
as in~(\ref{eq:representation}). This works since the ordering $<$ was constructed so that $\calR$
extends $\calR'$.
\end{proof}

\heading{Case I.2: No extended class corresponds to two maximal subwords.}\\
We number the classes according to their appearance in $\tau'$, i.e., $\Psi_i$ corresponds to the
subword $\tau'_i$. By Proposition~\ref{prop:structure}, we know that in any representation $\calR$
of $G$ the class $\Psi_i$ corresponds to two subwords $\tau_i$ and $\hat\tau_i$. The difficulty here
arises from the potential for $\tau'_i$ to be a subsequence of $\tau_i\hat\tau_i$, but of neither
$\tau_i$, nor $\hat\tau_i$. Figure~\ref{fig:caseI}(b) shows two potential extending representations.

\begin{figure}[b!]
\centering
\includegraphics{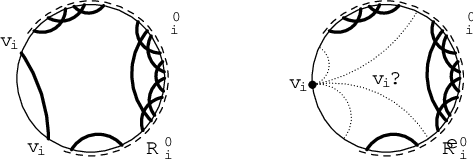}
\caption{The partial representation $\widetilde\calR'_i$ (with only a single endpoint of $v_i$
pre-drawn, depicted by a dot) is less restrictive with respect to the position of $v_i$. Therefore
it might be extendible even when $\calR'_i$ is not.}
\label{fig:two_representations}
\end{figure}

We solve this as follows. Instead of constructing just one partial representation $\calR'_i$ of
$G_i$ corresponding to the circular word $\tau'_i v_i v_i$, we construct an additional partial
representation $\widetilde\calR'_i$ corresponding to the circular word $\tau'_i v_i$, i.e., $v_i$
has only one endpoint pre-drawn.  Figure~\ref{fig:two_representations} shows that
$\widetilde\calR'_i$ is less restrictive: if $\calR'_i$ is extendible, then $\widetilde\calR'_i$ is
also extendible, but it might not be true the other way. For instance, every long chord in $\Phi_i$
alternates with $v_i$, so if some long chord has both endpoints pre-drawn in $\tau'_i$, 
$\calR'_i$ is necessarily non-extendible, but $\widetilde\calR'_i$ might be extendible.

The following lemma is the main trick of the algorithm and is essential to prove that it has cubic running
time. It states that, if $\tau'$ is extendible, at most one class can be forced to use $\widetilde\calR'_i$.

\begin{lemma} \label{lem:two_rep_cases}
In Case I.2, the representation $\calR'$ is extendible if and only if $\widetilde\calR'_i$ is
extendible for some $i$ and $\calR'_j$ is extendible for all $j \ne i$.
\end{lemma}

\begin{proof}
When $\calR_j$ corresponding to a word $v_j \tau_j v_j \hat\tau_j$ is an extension of $\calR'_j$ for
$j \ne i$, then $\tau'_j$ is a subsequence of, say, $\tau_j$.  On the other hand, when $\calR_i$
corresponding to a word $v_i \tau_i v_i \hat\tau_i$ is an extension of $\widetilde\calR'_i$, then
$\tau'_i$ is a subsequence of $\tau_i\hat\tau_i$, but might not be of $\tau_i$ or $\hat\tau_i$. We
use the circular ordering $\Psi_{i+1},\dots,\Psi_\ell,\Psi_1,\dots,\Psi_i$ of the classes and we
construct the representation $\calR$ as in~(\ref{eq:representation}): 
$$\boldsymbol{\tau_{i+1} \cdots \tau_\ell \tau_1 \cdots \tau_i}
\hat\tau_{i+1} \cdots \hat\tau_\ell \hat\tau_1 \cdots \hat\tau_{i-1}\boldsymbol{\hat\tau_i},$$
where all pre-drawn endpoints of $\tau'$ appear in those words written in bold.  It is easy to see
that $\calR$ extends $\calR'$ since $\tau'$ has no pre-drawn endpoints in $\hat\tau_{i+1} \cdots
\hat\tau_\ell \hat\tau_1 \cdots \hat\tau_{i-1}$.

For the other implication, suppose that $\calR$ extends $\calR'$. For contradiction, suppose that
two distinct partial representations $\calR'_i$ and $\calR'_j$ are not extendible. According to
Proposition~\ref{prop:structure}, the representation $\calR$ gives a representation $\calR_i$
corresponding to $v_i \tau_i v_i \hat\tau_i$ of $G_i$ and $\calR_j$ corresponding to $v_j \tau_j v_j
\hat\tau_j$ of $G_j$. Since $\calR'_i$ and $\calR'_j$ are non-extendible, we have that $\tau'_i$ is
neither a subsequence of $\tau_i$, nor $\hat\tau_i$, and similarly $\tau'_j$ is neither of $\tau_j$,
nor $\hat\tau_j$. Therefore, either $\tau_i\tau_j\hat\tau_i\hat\tau_j$, or
$\tau_j\tau_i\hat\tau_j\hat\tau_i$ is a subsequence of $\tau$, and we get that two maximal subwords
in $\tau'$ correspond to both $\Psi_i$ and $\Psi_j$ which is a contradiction.
\end{proof}

\begin{algorithm}[t!]
\centering
\begin{tabbing}
1.\quad Test whether each of $\calR'_2,\dots,\calR'_\ell$ is extendible. \\
2.\quad \textbf{If} \= two of $\calR'_2,\dots,\calR'_\ell$ are not extendible \textbf{then} {\sffamily REJECT}. \\
3.\quad \textbf{If} exactly one of $\calR'_2,\dots,\calR'_\ell$, denoted by $\calR'_i$, is not extendible \textbf{then} \\
4.\quad \> \textbf{If} $\widetilde\calR'_i$ and $\calR'_1$ are extendible \textbf{then}
{\sffamily ACCEPT} \textbf{else} {\sffamily REJECT}. \\
5.\quad \textbf{Else} (all of $\calR'_2,\dots,\calR'_\ell$ are extendible) \\
6.\quad \> \textbf{If} $\widetilde\calR'_1$ is extendible \textbf{then} {\sffamily ACCEPT}
\textbf{else} {\sffamily REJECT}. \\
\end{tabbing}
\vskip -1.2em
\caption{The subroutine for Case I.2.}
\label{alg:caseI2}
\end{algorithm}

Let $n = |V(G)|$ and let $\Psi_1$ be the largest class, so $|\Psi_i| \le n/2$ for $i > 1$.  If we want
to recursively test for each $\Psi_i$ whether both $\calR'_i$ and $\widetilde\calR'_i$ are
extendible, the running time might be exponential since we might have $|\Psi_1| \approx n$.
Fortunately, using Lemma~\ref{lem:two_rep_cases}, it is sufficient to test only one of $\calR'_1$
and $\widetilde\calR'_1$. We recursively test whether $\calR'_2,\dots,\calR'_\ell$ are extendible;
see the pseudocode of Algorithm~\ref{alg:caseI2}:
\begin{packed_itemize}
\item \emph{Two or more of $\calR'_2,\dots,\calR'_\ell$ are not extendible.} By
Lemma~\ref{lem:two_rep_cases}, $\calR'$ is non-extendible, the algorithm stops and outputs ``no''. 
\item \emph{Exactly one of $\calR'_2,\dots,\calR'_\ell$ is not extendible.} 
Let $\calR'_i$ be the non-extendible representation. We test whether $\widetilde\calR'_i$ and
$\calR'_1$ are extendible. If at least one is non-extendible, the algorithm stops and outputs
``no''. If both are extendible, we similarly join in $\calR$ the representations
$\calR_1,\dots,\calR_\ell$ according to~(\ref{eq:representation}) as described in the proof of
Lemma~\ref{lem:two_rep_cases}.
\item \emph{All representations $\calR'_2,\dots,\calR'_\ell$ are extendible.} We have
representations $\calR_2,\dots,\calR_\ell$ where $\calR_i$ extends $\calR'_i$.  We test whether the
partial representation $\widetilde\calR'_1$ is extendible. If not, the algorithm stops and outputs
``no''. If it extends, we get a representation $\calR_1$ of $\widetilde G_1$. We construct the
representation $\calR$ using~(\ref{eq:representation}) as described in the proof of
Lemma~\ref{lem:two_rep_cases}.
\end{packed_itemize}

\begin{lemma} \label{lem:correctness_of_caseI2}
In Case I.2, the representation $\calR'$ is extendible if and only if the algorithm constructs it.
\end{lemma}

\begin{proof}
We know that $\widetilde\calR'_i$ is extendible when $\calR'_i$ is extendible.
Lemma~\ref{lem:two_rep_cases} states that $\calR'$ is extendible if and only if at most one of
$\calR'_i$ is non-extendible while $\widetilde\calR'_i$ is extendible. The algorithm tests this in
Case I.2, while postponing $\Psi_1$ until it knows which of $\calR'_1$ and $\widetilde\calR'_1$ needs
to be tested.
\end{proof}

\subsection{Case II: A Trivial Maximal Split Between $\boldsymbol A$ and $\boldsymbol B$}
\label{sec:caseII}

Let $A = \{a\}$ and $\short(A) = \emptyset$. In Section~\ref{sec:trivial_maximal_splits} we 
characterized all possible representations $\calR$ in terms of representations of connected 
components $C$ of $G \setminus a$. 
We just need to test whether one of them is compatible with the partial representation $\calR'$
corresponding to the circular word $\tau'$. Similarly as in Section~\ref{sec:caseI}, we may assume
that every connected component $C$ has at least one endpoint in $\tau'$; otherwise, we can deal with
it trivially.

\heading{Case II.1: Both endpoints of $\boldsymbol{a}$ appear in $\boldsymbol\tau'$.}
The circular word $\tau'$ is composed of $k$ and $k'$ \emph{maximal subwords} $\tau' = a
\tau'_1\tau'_2\cdots\tau'_k a \hat\tau'_{k'} \hat\tau'_{k'-1} \cdots \hat\tau'_1$ such that each
$\tau'_i$ contains only symbols of one connected component $C$ and similarly for each $\hat\tau'_i$.
According to Proposition~\ref{prop:art_structure}, each connected component $C$ corresponds to at
most two different maximal subwords. If a connected component $C$ corresponds to two subwords
$\tau'_i$ and $\hat\tau'_j$, then $a\tau'_i a \hat\tau'_j$ is a subsequence of $\tau'$.
Also, if two components $C$ and $\hat C$ each correspond to two different maximal subwords, then
occurrences of these subwords do not alternate in $\tau'$.  Otherwise we reject the input.

\begin{figure}[t!]
\centering
\includegraphics{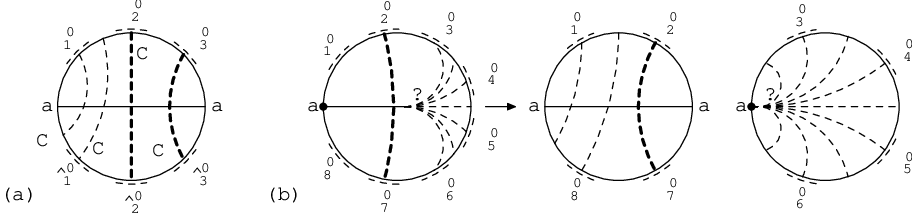}
\caption[]{Each dashed line represents one connected component. (a) An example of Case II.1. We have
$\tau' = a\tau'_1\tau'_2\tau'_3a\hat\tau'_3\hat\tau'_2\hat\tau'_1$ and four connected components
$C_\alpha$ (corresponding to $\tau'_1$), $\Psi_\beta$ (to $\tau'_2$ and $\hat\tau'_2$), $\Psi_\gamma$
(to $\tau'_3$ and $\hat\tau'_3$), and $\Psi_\delta$ (to $\hat\tau'_1$). We get that $C_\alpha <
C_\beta < C_\gamma$ and $C_\delta < C_\beta < C_\gamma$, so one possible linear ordering is
$C_\alpha,C_\delta,C_\beta,C_\gamma$, and $\alpha=1$, $\delta=2$, $\beta=3$, and $\gamma=4$.
By Lemma~\ref{lem:correctness_of_caseII1}, $\calR'$ is extendible if and only if
$\calR'_1,\dots,\calR'_4$ are extendible.
(b) An example of Case II.2. On the left, we have a connected component corresponding to two maximal
subwords $\tau'_2$ and $\tau'_7$. Therefore, every extending representation has the subsequence
$a\tau'_2a\tau'_7$. We divide the problem into two depicted subproblems, one of Case II.1, the other
of Case II.2 with each component $C_i$ corresponding to exactly one maximal subword $\tau'_i$.}
\label{fig:caseII}
\end{figure}

Next, we find a linear ordering of $\ell$ connected components as follows. We order $C < C'$ if $C$
corresponds to a subword $\tau'_s$ and $C'$ to a subword $\tau'_t$ for $s < t$, or $C$ to
$\hat\tau'_s$ and $C'$ to $\hat\tau'_t$ for $s < t$. We obtain a linear ordering $C_1,\dots,C_\ell$
as any linear extension. Since subwords of all connected components with two subwords in $\tau'$ do
not alternate, we get that $<$ is acyclic and a linear extension always exists.  Suppose that we
renumber the maximal subwords of $\tau'$ in such a way that $C_i$ corresponds to $\tau'_i$ and
$\hat\tau'_i$ (one of them possibly empty). Let $G_i$ be the subgraph of $G$ induced by $V(C_i) \cup
\{a\}$. Let $\calR'_i$ be the partial representation of $G_i$ corresponding to the circular word $a
\tau'_i a \hat\tau'_i$, so $\tau' = a \tau'_1 \cdots \tau'_\ell a \hat\tau'_1 \cdots \hat\tau'_\ell$.
Figure~\ref{fig:caseII}(a) shows an example.

\begin{lemma} \label{lem:correctness_of_caseII1}
In Case II.1, the representation $\calR'$ is extendible if and only if the representations
$\calR'_1,\dots,\calR'_\ell$ of the graphs $G_1,\dots,G_\ell$ are extendible.
\end{lemma}

\begin{proof}
Suppose that $\calR$ extends $\calR'$. According to Proposition~\ref{prop:art_structure}, the
representations of $C_1,\dots,C_\ell$ are ordered along the circle, and so we obtain
representations $\calR_1,\dots,\calR_\ell$ extending $\calR'_1,\dots,\calR'_\ell$.

For the other implication, we just take $\calR_1,\dots,\calR_\ell$ and combine them to form $\calR$ 
as in~(\ref{eq:art_representation}). This works since the ordering $<$ was constructed so that
$\calR$ extends $\calR'$.
\end{proof}

\heading{Case II.2: A single endpoint of $\boldsymbol{a}$ appears in $\boldsymbol\tau'$.}
The circular word $\tau'$ is composed of $k$ \emph{maximal subwords} $\tau' = a \tau'_1 \tau'_2
\cdots \tau'_k$ such that each $\tau'_i$ contains only symbols of one connected component $C$.
According to Proposition~\ref{prop:art_structure}, each connected component $C$ corresponds to at
most two different maximal subwords. Also, if two components $C$ and $\hat C$ each correspond to two
different maximal subwords, then occurrences of these subwords do not alternate in $\tau'$.
Otherwise we reject the input.

Suppose there is a component $C$ corresponding to two maximal subwords $\tau'_s$ and $\tau'_t$ for
$s < t$. Further let $C$ be such a component that maximizes the value $s$. 
In every extending representation, we have the subsequence
$a \tau'_s a \tau'_t$, so we can assume that the second endpoint of $a$ is pre-drawn in between
$\tau'_s$ and $\tau'_t$. We divide testing whether $\calR'$ is extendible into two subproblems.
We deal with the connected components of the circular word $a \tau'_1 \tau'_2 \cdots \tau'_s a \tau'_t
\tau'_{t+1} \cdots \tau'_k$ exactly as in Case II.1. It remains to decide whether $a \tau'_{s+1}
\cdots \tau'_{t-1}$ is extendible where each connected component corresponds to precisely one
maximal subword (note: when no such component $C$ exists, we have precisely this situation). 
Figure~\ref{fig:caseII}(b) shows an example. 

Suppose that we rename $\tau' = a \tau'_1 \cdots \tau'_\ell$ such that $\tau'_i$ corresponds to the
connected component $C_i$. Similarly to Case I.2, the difficulty comes from the fact that some
$\tau'_i$ might be a subsequence of $\tau_i\hat\tau_i$ of~(\ref{eq:art_representation}) in an
extending representation, but not of $\tau_i$ or $\hat\tau_i$. We consider two partial
representations for each $G_i$: the partial representation $\calR'_i$ corresponding to $a \tau'_i a$
and $\widetilde\calR'_i$ corresponding to $a \tau'_i$. Again, if $\calR'_i$ is extendible, then
$\widetilde\calR'_i$ is also extendible.

\begin{lemma} \label{lem:caseII_two_rep_cases}
In Case II.2 with no connected component correspond to two maximal subwords of $\tau'$, the
representation $\calR'$ is extendible if and only if $\widetilde\calR'_i$ is extendible for some $i$
and $\calR'_j$ is extendible for all $j \ne i$.
\end{lemma}

\begin{proof}
When $\calR_j$ corresponding to a word $a \tau_j a \hat\tau_j$ is an extension of $\calR'_j$ for $j
\ne i$, then $\tau'_j$ is a subsequence of, say, $\tau_j$ for $j < i$ and of $\hat\tau_j$
for $j > i$.  On the other hand, when $\calR_i$ corresponding to a word $a \tau_i a \hat\tau_i$ is
an extension of $\widetilde\calR'_i$, then $\tau'_i$ is a subsequence of $\tau_i\hat\tau_i$, but
might not be of $\tau_i$ or $\hat\tau_i$.  We use the linear ordering $C_1, \dots, C_{i-1}, C_\ell,
C_{\ell-1}, \dots, C_i$ of the connected components and we construct the representation $\calR$ as
in~(\ref{eq:art_representation}): 
$$\boldsymbol{a \tau_1 \cdots \tau_{i-1}} \tau_\ell \cdots \tau_{i+1} \boldsymbol{\tau_i} a
\boldsymbol{\hat\tau_i \cdots \hat\tau_\ell} \hat\tau_{i-1} \cdots \hat\tau_1.$$
where all pre-drawn endpoints of $\tau'$ appear in those words written in bold.  It is easy to see
that $\calR$ extends $\calR'$ since there are no pre-drawn endpoints in $\tau_\ell \cdots
\tau_{i+1}$ and in $\hat\tau_{i-1} \cdots \hat\tau_1$.

For the other implication, suppose that $\calR$ corresponding to $\tau$ extends $\calR'$, and we add
into $\tau'$ the position of the other endpoint of $a$. It splits at most one maximal word
$\tau'_i$, so $a \tau'_j a$ is a subsequence of $\tau$ and $\calR'_j$ is extendible. Since $a
\tau'_i$ is a subsequence of $\tau$, we get that $\widetilde\calR'_i$ is extendible.
\end{proof}

The rest of this case proceeds exactly as Case I.2.

\begin{lemma} \label{lem:correctness_of_caseII2}
In Case II.2, the representation $\calR'$ is extendible if and only if the algorithm constructs it.
\end{lemma}

\begin{proof}
The proof is similar to Lemma~\ref{lem:correctness_of_caseI2}.
\end{proof}

\heading{Case II.3: No endpoint of $\boldsymbol{a}$ appears in $\boldsymbol\tau'$.}
As in Case II.2, the circular word $\tau'$ is composed of $k$ \emph{maximal subwords} $\tau' =
\tau'_1 \tau'_2 \cdots \tau'_k$. If two components $C$ and $\hat C$ each correspond to two different
maximal subwords, then occurrences of these subwords do not alternate in $\tau'$.  Otherwise we
reject the input. Also, if some connected component $C$ corresponds to two subwords $\tau'_i$ and
$\tau'_j$, then $a \tau'_i a \tau'_j$ is a subsequence of every extending representation. Therefore,
existence of such a component restricts the possible positions of endpoints of $a$, so we divide
this case into two subcases.

\begin{figure}[b!]
\centering
\includegraphics{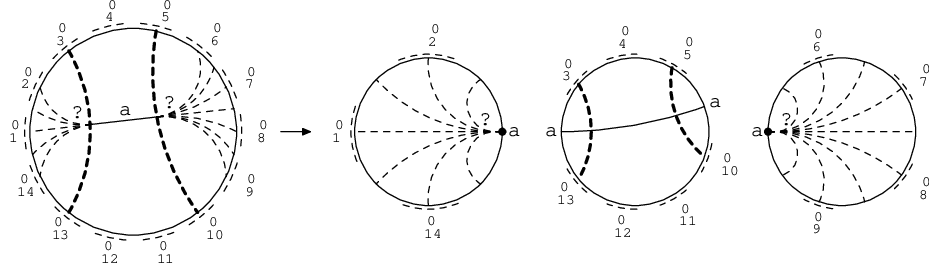}
\caption{An example of Case II.3a. On the left, we have two connected components corresponding to
two maximal subwords $\tau'_3$ and $\tau'_{13}$, and $\tau'_5$ and $\tau'_{10}$. We put $p = 3$,
$q=13$, $s = 5$, and $t = 10$. We divide testing whether $\calR'$ is extendible into three
subproblems depicted on the right.}
\label{fig:caseII3a}
\end{figure}

\heading{Case II.3a: Some component has two maximal subwords in $\boldsymbol\tau'$.}
By a suitable renaming of the subwords, let $C$ be the connected component corresponding to
$\tau'_p$ and $\tau'_q$ such that $p < q$, $p$ is minimal, and
$\tau'_{q+1},\dots,\tau'_{\ell},\tau'_1,\dots,\tau'_{p-1}$ correspond to connected components having
only one maximal subword in $\tau'$. Similarly, let $C'$ be the connected component corresponding to
$\tau'_s$ and $\tau'_t$ such that $s < t$ and all $\tau'_{s+1},\dots,\tau'_{t-1}$ correspond to
connected components having only one maximal subword in $\tau'$, and possibly $C = C'$. If $\calR'$
is extendible, we get that every connected component corresponding to two maximal subwords $\tau'_x$
and $\tau'_y$ has $p \le x \le s < t \le y \le q$; otherwise we reject the input.
Figure~\ref{fig:caseII3a} shows an example.

It follows that every extending representation has $a \tau'_p \tau'_s a \tau'_t \tau'_q$ as a
subsequence. Similarly as Case II.2, we can divide testing whether $\calR'$ is extendible into three
subproblems:
\begin{packed_itemize}
\item Testing using Case II.2 whether the partial representation $\tau'_{q+1} \cdots \tau'_{\ell}
\tau'_1 \cdots \tau'_{p-1}a$ is extendible.
\item Testing using Case II.1 whether the partial representation $a \tau'_p \cdots \tau'_s a \tau'_t
\cdots \tau'_q$ is extendible.
\item Testing using Case II.2 whether the partial representation $a \tau'_{s+1} \cdots \tau'_{t-1}$
is extendible.
\end{packed_itemize}

\begin{lemma} \label{lem:correctness_of_caseII3a}
In Case II.3a, the representation $\calR'$ is extendible if and only if the algorithm constructs it.
\end{lemma}

\begin{proof}
This is implied by Lemmas~\ref{lem:correctness_of_caseII1} and~\ref{lem:correctness_of_caseII2}.
\end{proof}

\heading{Case II.3b: No component has two maximal subwords in $\boldsymbol\tau'$.}
Let $\tau'_i$ correspond to the connected component $C_i$, and define $\calR'_i$ and
$\widetilde\calR'_i$ exactly as in Case II.2. Similarly to Case II.2, the difficulty comes from
the fact that some $\tau'_i$ might correspond to both $\tau_i$ and $\hat\tau_i$
of~(\ref{eq:art_representation}) in an extending representation. Since we are placing two endpoints
of $a$, we might have two such components $C_i$ and $C_j$.

\begin{lemma} \label{lem:caseII3_two_rep_cases}
The representation $\calR'$ is extendible if and only if $\widetilde\calR'_i$ and
$\widetilde\calR'_j$ are extendible for some $i$ and $j$, and $\calR'_k$ is extendible for all
$k \ne i,j$.
\end{lemma}

\begin{proof}
Let $i < j$. When $\calR_k$ corresponding to a word $a \tau_k a \hat\tau_k$ is an extension of
$\calR'_k$ for $k \ne i,j$, then $\tau'_k$ is a subsequence of, say, $\tau_k$ for $k < j$ and of
$\hat\tau_k$ for $k > j$.  On the other hand, when $\calR_i$ corresponding to a word $a \tau_i a
\hat\tau_i$ is an extension of $\widetilde\calR'_i$, then $\tau'_i$ is a subsequence of
$\tau_i\hat\tau_i$, but might not be of $\tau_i$ or $\hat\tau_i$, and similarly for $\calR_j$. We
use the linear ordering $C_i, C_{i-1}, \dots, C_1, C_{i+1}, \dots, C_{j-1}, C_{\ell}, C_{\ell-1},
\dots, C_{j+1}, C_j$ of the connected components and we construct the representation $\calR$ as
in~(\ref{eq:art_representation}): 
$$a \boldsymbol{\tau_i} \hat\tau_{i-1} \cdots \hat\tau_1 \boldsymbol{\tau_{i+1} \cdots \tau_{j-1}}
\tau_{\ell} \cdots \tau_{j+1} \boldsymbol{\tau_j} a \boldsymbol{\hat\tau_j \cdots \hat\tau_\ell}
\hat\tau_{j-1} \cdots \hat\tau_{i+1} \boldsymbol{\tau_1 \cdots \tau_{i-1} \hat\tau_i},$$
where all pre-drawn endpoints of $\tau'$ appear in those words written in bold.  It is easy to see
that $\calR$ extends $\calR'$ since there are no pre-drawn endpoints in $\hat\tau_{i-1} \cdots
\hat\tau_1$, in $\tau_{\ell} \cdots \tau_{j+1}$, and in $\hat\tau_{j-1} \cdots \hat\tau_{i+1}$.

For the other implication, suppose that $\calR$ corresponding to $\tau$ extends $\calR'$, and we add
into $\tau'$ the positions of the endpoints of $a$. It is not possible that both endpoints split the
same maximal word $\tau'_i$, otherwise the remaining components $C_k$ would alternate with $C_i$. 
It is additionally not possible that two maximal words are split by the same endpoint. 
So at most two maximal words $\tau'_i$ and $\tau'_j$ are split by the endpoints of $a$.
Therefore, for every $k \ne i,j$, we have $a \tau'_k a$ as a subsequence of $\tau$, so $\calR'_k$ is
extendible.  Since $a \tau'_i$ and $a \tau'_j$ are subsequences of $\tau$, we get that
$\widetilde\calR'_i$ and $\widetilde\calR'_j$ are also extendible.
\end{proof}

Let $n = |V(G)|$ and let $C_1$ be the largest component, so $|V(C_i)| \le n/2$ for $i > 1$.
The algorithm works similarly to Case I.2; see Algorithm~\ref{alg:caseII3} for a pseudocode.
So we test the extendibility of only one of $\calR'_1$ and $\widetilde\calR'_1$ while testing both 
types of representations for at most two other graphs $G_i$ and $G_j$.

\begin{algorithm}[t!]
\centering
\begin{tabbing}
1.\quad Test whether each of $\calR'_2,\dots,\calR'_\ell$ is extendible. \\
2.\quad \textbf{If} \= three of $\calR'_2,\dots,\calR'_\ell$ are not extendible \textbf{then} {\sffamily REJECT}. \\
3.\quad \textbf{If} exactly two of $\calR'_2,\dots,\calR'_\ell$, denoted $\calR'_i$ and $\calR'_j$,
are not extendible \textbf{then} \\
4.\quad \> \textbf{If} $\widetilde\calR'_i$, $\widetilde\calR'_j$ and $\calR'_1$ are extendible \textbf{then}
{\sffamily ACCEPT} \textbf{else} {\sffamily REJECT}. \\
5.\quad \textbf{If} exactly one of $\calR'_2,\dots,\calR'_\ell$, denoted $\calR'_i$, is not extendible \textbf{then} \\
6.\quad \> \textbf{If} $\widetilde\calR'_i$ and $\widetilde\calR'_1$ are extendible \textbf{then}
{\sffamily ACCEPT} \textbf{else} {\sffamily REJECT}. \\
7.\quad \textbf{Else} (all of $\calR'_2,\dots,\calR'_\ell$ are extendible) \\
8.\quad \> \textbf{If} $\widetilde\calR'_1$ is extendible \textbf{then} {\sffamily ACCEPT}
\textbf{else} {\sffamily REJECT}. \\
\end{tabbing}
\vskip -1.2em
\caption{The subroutine for Case II.3b.}
\label{alg:caseII3}
\end{algorithm}

\begin{lemma} \label{lem:correctness_of_caseII3b}
In Case II.3b, the representation $\calR'$ is extendible if and only if the algorithm constructs it.
\end{lemma}

\begin{proof}
We use Lemma~\ref{lem:caseII3_two_rep_cases} similarly as in the proof of
Lemmas~\ref{lem:correctness_of_caseI2} and~\ref{lem:correctness_of_caseII2}.
\end{proof}

\subsection{Analysis of the Algorithm}
\label{sec:analysis}

By using the established results, we show that the partial representation extension problem of
circle graphs can be solved in cubic time.

\begin{lemma} \label{lem:correctness}
The described algorithm correctly decides whether the partial representation $\calR'$ of $G$ is
extendible.
\end{lemma}

\begin{proof}
If the input graph $G$ is prime, we just test both representations whether they extend $\tau'$.
If the input graph $G$ contains a non-trivial split, we modify it into a maximal split between $A$ and
$B$ using Lemma~\ref{lem:maximal_split}. Next, we proceed by Case I or Case II, depending whether
the maximal split is trivial or not.  For Case I, the algorithm is correct by
Lemmas~\ref{lem:correctness_of_caseI1} and~\ref{lem:correctness_of_caseI2}. For Case II, the
algorithm is correct by Lemmas~\ref{lem:correctness_of_caseII1}, \ref{lem:correctness_of_caseII2},
\ref{lem:correctness_of_caseII3a}, and~\ref{lem:correctness_of_caseII3b}.
\end{proof}

\begin{lemma} \label{lem:complexity}
The running time of the algorithm is $\O(n^3)$ where $n$ is the number of vertices.
\end{lemma}

\begin{proof}
Let $T(n)$ denote the time complexity of the algorithm for at most $n$ vertices in the worst case.
We want to show that $T(n) = \O(n^3)$.

As described, we can test whether the graph $G$ is prime and construct a unique representation
$\tau$ in quasilinear time using~\cite{GPTC13}, but for the purpose of our analysis $\O(n^2)$ is
sufficient.  Since each symbol appears twice in $\tau$, we can easily test in linear time whether
$\tau'$ is a subsequence of $\tau$ or its reversal.  If $G$ is not prime, then we can find a
non-trivial split between $A'$ and $B'$ using~\cite{dalhaus} and modify it using
Lemma~\ref{lem:maximal_split} into a maximal split between $A$ and $B$ such that $A' \subseteq A$
and $B' \subseteq B$. Both can be achieved in linear time.

\emph{Case I.}
We compute the $\sim$ relation in time $\O(n^2)$. 
\begin{packed_itemize}
\item In Case I.1, we divide the problem into $\ell$ smaller disjoint subproblems of total size $n$,
each of size $n_i+1$ solvable by induction hypothesis in time $\O(n_i^3)$, so the total running time is
$\O(n^3)$.
\item In Case I.2, we test both representations $\calR'_i$ and $\widetilde\calR'_i$ for at most one
extended class of size $|\Psi_i| \le {n \over 2}$, while we test exactly one of these
representations of all remaining extended classes. We get the following recursion:
$$T(n) \le T(n/2+1) + \sum_j T(|\Psi_j|+1) + \O(n^2) \le T(n/2+1) + \O(n^3).$$
By the Master Theorem, we get that $T(n) \le \O(n^3)$. Since the depth of the recursion is at most
linear, each level of the recursion adds to at most $\O(n^2)$ and we get $\O(n^3)$ in total over all
levels.
\end{packed_itemize}

\emph{Case II.}
We find connected components of $G \setminus a$ in linear time.
\begin{packed_itemize}
\item In Case II.1, the analysis is similar as in Case I.1.
\item In Case II.2, we divide the input into two disjoint subproblems, one is solved as in Case
II.1, the other as in Case I.2. Therefore, the total running time is $\O(n^3)$.
\item In Case II.3a, we divide the input into three disjoint subproblems solved using Case II.1 and
Case II.2, so the total running time is $\O(n^3)$.
\item In Case II.3b, we test both representations $\calR'_i$ and $\widetilde\calR'_i$ for at most two
extended class of size $|\Psi_i| \le {n \over 2}$, while we test exactly one of these
representations of all remaining extended classes. We get the following recursion:
$$T(n) \le 2T(n/2+1) + \sum_j T(|\Psi_j|+1) + \O(n^2) \le 2T(n/2+1) + \O(n^3).$$
By the Master Theorem, we again get that $T(n) \le \O(n^3)$.
\end{packed_itemize}

Therefore, the total running time is $\O(n^3)$.
\end{proof}

The proof of the main result in this paper now follows easily.

\begin{proof}[Proof of Theorem~\ref{thm:ext_circle}]
The result is implied by Lemma~\ref{lem:correctness} and Lemma~\ref{lem:complexity}.
\end{proof}

\section{Simultaneous Representations of Circle Graphs} \label{sec:simultaneous_representations}

In this section, we give two results concerning the simultaneous representation problem for circle graphs: We
show that this problem is \cNP-complete and \cFPT\ in the size of the common intersection.
Formally, we deal with the following decision problem:

\computationproblem
{Simultaneous Representation for Circle Graphs -- $\simrep(\circle)$}
{Graphs $G_1,\dots, G_k$ such that $G_i \cap G_j = I$ for all $i \ne j$.}
{Do there exist representations $\calR_1,\dots,\calR_k$ of $G_1,\dots, G_k$ which use the same representation of the vertices of $I$?}
{10.7cm}

\begin{proof}[Proof of Theorem~\ref{prop:simrep_npc}]
To show that $\simrep(\circle)$ is \cNP-complete, we reduce it from the \emph{total ordering
problem}:

\computationproblem
{The total ordering problem - \totalorder}
{A finite set $S$ and a finite set $T$ of triples from $S$.}
{Does there exist a total ordering $<$ of $S$ such that for all $(x, y, z) \in T$ either $x < y <
z$, or $z < y < x$?}
{8.6cm}

\noindent Opatrny~\cite{opatrny} proved this problem is \cNP-complete.

\begin{figure}[b!]
\centering
\includegraphics{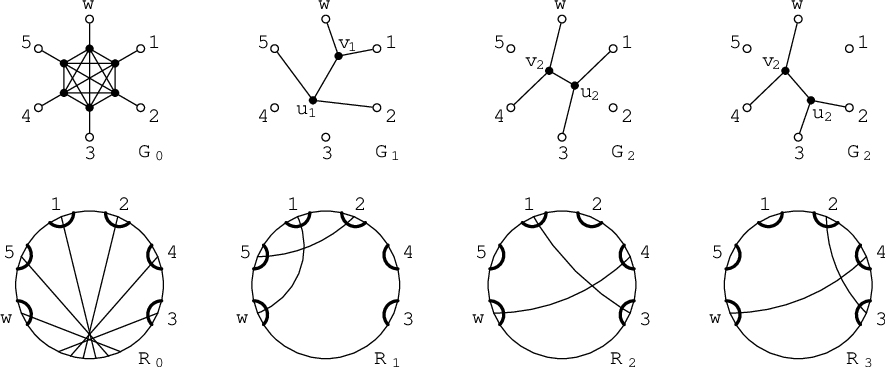}
\caption{Let $S = \{1,2,3,4,5\}$ and $T$ consisting of three triples $(5,1,2)$, $(1,4,3)$ and
$(2,4,3)$ be the instance of \totalorder. We construct graphs $G_0,\dots,G_3$ depicted in the top,
with the common vertices $I$ depicted in white.  Possible simultaneous representations are depicted
in the bottom, giving the total ordering $5 < 1 < 2 < 4 < 3$.}
\label{fig:simrep_example}
\end{figure}

Given an instance $(S,T)$ of \totalorder\ and let $s = |S|$ and $t = |T|$. We construct a set of
$t+1$ graphs $G_0,G_1, \ldots ,G_t$ as follows, so the number $k$ from $\simrep(\circle)$ is equal
$t+1$. The intersection of $G_0,G_1, \ldots ,G_t$ is an independent set $I=S\cup \{w\}$ where $w$ is
a special vertex. The graph $G_0$ consists of a clique $K_{s+1}$, and to each vertex of this clique
we attach exactly one vertex of $I$ as a leaf. The graph $G_i$ corresponds to the $i$-th constraint
$(x_i, y_i, z_i) \in T$. In addition to $I$, each $G_i$ contains two vertices $u_i$ and $v_i$ of
degree three, such that $u_i$ is adjacent to $v_i$, $x_i$ and $z_i$, and $v_i$ is further adjacent
to $y_i$ and the special vertex $w$. See Fig.~\ref{fig:simrep_example} for an example of this
construction.

The clique in $G_0$ defines a split where each class of $\sim$ is a singleton. According to
Proposition~\ref{prop:structure}, every representation $\calR_0$ of $G_0$ places the elements of $I$
in some circular ordering $wws_1s_1s_2s_2\cdots s_ss_s$ which corresponds to the total ordering $s_1
< s_2 < \cdots < s_s$.  Now the representations $\calR_1,\dots,\calR_t$ of $G_1,\dots,G_t$ can be
constructed if and only if all the total ordering constraints are satisfied. This implies that there
exists a solution $\calR_0,\dots,\calR_t$ of $G_0,\dots,G_t$ if and only if the instance $(S,T)$ of
\totalorder\ is solvable.
\end{proof}

Further, we show that the problem is \cFPT\ in size of the common subgraph $I$.

\begin{proof}[Proof of Corollary\ref{cor:simrep_fpt}]
We just consider all possible representations of the common subgraph $I$ which are all words of
length $2|V(I)|$. Each word gives some partial representation $\calR'$. We just solve $k$ instance
of $\ext(\circle)$ for each $G_i$ and the partial representation $\calR'$ of $I$, which can be done
in polynomial time according to Theorem~\ref{thm:ext_circle}.
\end{proof}

\section{Conclusions} \label{sec:conclusions}

The structural results described in Section~\ref{sec:structural_results}, namely
Propositions~\ref{prop:structure} and~\ref{prop:art_structure}, are the main new tools developed in
this paper. Using it, one can easily work with the structure of all representations which is a key
component of the algorithm of Section~\ref{sec:algorithm} that solves the partial representation
extension problem for circle graphs. The algorithm works with the recursive structure of all
representations and matches the partial representation on it. Proposition~\ref{prop:structure} also
seems to be useful in attacking the following open problems:

\begin{question}
What is the complexity of $\simrep(\circle)$ for a fixed number $k$ of graphs? In particular, what
is it for $k=2$?
\end{question}

Recall that in the bounded representation problem, we give for some chords two circular arcs and we
want to construct a representation which places endpoints into these circular arcs.

\begin{question}
What is the complexity of the bounded representation problem for circle graphs? This question is also
 open for interval graphs and proper interval graphs.
\end{question}

\heading{Permutation Graphs.}
Permutation graphs are intersection graphs of segments between two parallel lines.  So every
permutation representation of $G$ consists of two words $\tau$ and $\hat\tau$, each containing each
vertex $V(G)$ exactly once, and $uv \in E(G)$ if and only if their order in $\tau$ and $\hat\tau$
differs. We denote the class by \perm. 

Let $\hat\tau_R$ be the reversal of $\hat\tau$.  Since $\hat\tau$ is a circle representation of $G$, it
follows that every permutation graph is a circle graph. More strongly, a graph $G$ is a permutation
graph if and only if $\widetilde G$ constructed from $G$ by adding a universal vertex $u$ is a
circle graph, since $u\tau u\hat\tau_R$ is a circle representation of $\widetilde G$.

The partial representation problem for permutation graphs is studied in~\cite{KKKW} and solved in
time $\O(n^3)$. The following results gives an alternative algorithm running in time $\O(n^3)$ as
well.

\begin{proposition}
The problem $\ext(\perm)$ reduces in time $\O(n+m)$ to $\ext(\circle)$. 
\end{proposition}

\begin{proof}
Let $G$ be a permutation graph with a partial representation $\calR'$ corresponding to two words
$\tau'$ and $\hat\tau'$. The problem $\ext(\perm)$ asks whether there exists words $\tau$ and
$\hat\tau$ representing $\calR$ such that $\tau'$ and $\hat\tau'$ are subsequences of $\tau$ and
$\hat\tau$, respectively. The reduction constructs the circle graph $\widetilde G$ by adding a
universal vertex $u$ to $G$ and the partial representation $\widetilde\calR'$ given by the circular
word $u\tau' u \hat\tau'_R$.  The reduction clearly works in linear time. It is correct since the
partial representation $\calR'$ of $G$ is extendible if and only if $\widetilde\calR'$ of
$\widetilde G$ is extendible.
\end{proof}

\heading{Minimal Split Decomposition and Split Trees.}
A split decomposition of $G$ works as follows.  Consider a split between $A$ and $B$. We replace $G$
by the graphs $G_A$ and $G_B$ defined in Section~\ref{sec:conditions_split}. Then we apply the
decomposition recursively on $G_A$ and $G_B$, and we stop on prime graphs containing no splits. We
note that by different orders of splits, different decompositions of $G$ may be constructed. A split
decomposition can be computed in linear time~\cite{dalhaus}.

A split decomposition is called \emph{minimal} if it is constructed by the least number of splits.
Suppose that we also stop on \emph{degenerate graphs} which are complete graphs $K_n$ and stars
$S_n = K_{1,n}$.  Cunningham~\cite[Theorem 3]{Cunningham82} proved that the minimal split decomposition of a connected
graph stopping on prime and degenerate graphs is unique.

\begin{figure}[t!]
\centering
\includegraphics{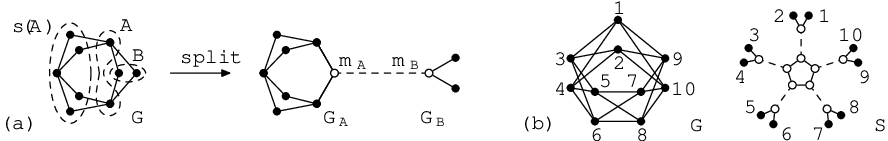}
\caption{(a) An example of a split of the graph $G$. The marker vertices are depicted in white.  The
tree edge is depicted by a dashed line.  (b) The split tree $S$ of the graph $G$.}
\label{fig:split_tree}
\end{figure}

The unique \emph{split tree} $S$ representing a graph $G$ encodes the minimal split
decomposition~\cite{GPTC13}. A split tree is a graph with two types of vertices (normal and marker
vertices) and two types of edges (normal and tree edges). We initially put $S = G$ and modify it
according to the minimal split decomposition. If the minimal decomposition contains a split between
$A$ and $B$ in $G$, then we replace $G$ in $S$ by the graphs $G_A$ and $G_B$, and connect the marker
vertices $m_A$ and $m_B$ by a \emph{tree edge} (see Fig.~\ref{fig:split_tree}a).  We repeat this
recursively on $G_A$ and $G_B$; see Fig.~\ref{fig:split_tree}b. Each prime and degenerate graph is a
\emph{node} of the split tree.  A node that is incident with exactly one tree edge is called a
\emph{leaf node}.

The minimal split decompositions and the split trees can be
computed in quasi-linear time~\cite{GPTC13}. Similarly as in Propositions~\ref{prop:structure}
and~\ref{prop:art_structure}, it should be possible to derive every circle representation of a
connected graph $G$ from the split tree $S$, but the precise statement is unclear.  It is a natural
question whether split trees can be used to solve the partial representation extension problem:

\begin{question}
Is it possible to use split trees $S$ to solve $\ext(\circle)$? Can it be done faster that in time
$\O(n^3)$?
\end{question}

\acknowledgment{
We want to thank to an anonymous reviewer for pointing out than the results described in
Section~\ref{sec:structural_results} work only for maximal splits.

The second author is supported by the People Programme (Marie Curie Actions) of the European Union's
Seventh Framework Programme (FP7/2007-2013) under REA grant agreement no [291734].

The third author is supported by CE-ITI (P202/12/G061 of GA\v{C}R) and Charles University as GAUK
1334217.}

\bibliographystyle{plain}
\articlebibliography{extending_circle_graphs}

\end{article}

\end{document}